\documentclass[twocolumn]{aastex63}
\usepackage{amsmath,natbib,color}
\usepackage{appendix}
\usepackage{supertabular}

\definecolor{red}{rgb}{1.0,0.0,0.0}

\received{October 25, 2019}
\revised{February 27, 2020}
\accepted{April 2, 2020}

\submitjournal{AJ}

\shorttitle{Baade's Window: an astrometric calibration field}
\shortauthors{Nguyen et al.}

\begin{document}

\title{HD 165054: an astrometric calibration field for high-contrast imagers in Baade's Window}

\author[0000-0002-9350-4763]{Meiji M. Nguyen}
\affiliation{Department of Astronomy, University of California, Berkeley, CA 94720, USA}

\author[0000-0002-4918-0247]{Robert J. De Rosa}
\affiliation{European Southern Observatory, Alonso de Cordova 3107, Vitacura, Santiago, Chile}
\affiliation{Kavli Institute for Particle Astrophysics and Cosmology, Stanford University, Stanford, CA 94305, USA}

\author[0000-0003-0774-6502]{Jason J. Wang}
\altaffiliation{51 Pegasi b Fellow}
\affiliation{Department of Astronomy, California Institute of Technology, Pasadena, CA 91125, USA}

\author[0000-0002-0792-3719]{Thomas M. Esposito}
\affiliation{Department of Astronomy, University of California, Berkeley, CA 94720, USA}

\author{Paul Kalas}
\affiliation{Department of Astronomy, University of California, Berkeley, CA 94720, USA}
\affiliation{SETI Institute, Carl Sagan Center, 189 Bernardo Ave.,  Mountain View CA 94043, USA}

\author{James R. Graham}
\affiliation{Department of Astronomy, University of California, Berkeley, CA 94720, USA}

\author[0000-0003-1212-7538]{Bruce Macintosh}
\affiliation{Kavli Institute for Particle Astrophysics and Cosmology, Stanford University, Stanford, CA 94305, USA}

\author[0000-0002-5407-2806]{Vanessa P. Bailey}
\affiliation{Jet Propulsion Laboratory, California Institute of Technology, Pasadena, CA 91109, USA}

\author[0000-0002-7129-3002]{Travis Barman}
\affiliation{Lunar and Planetary Laboratory, University of Arizona, Tucson AZ 85721, USA}

\author{Joanna Bulger}
\affiliation{Institute for Astronomy, University of Hawaii, 2680 Woodlawn Drive, Honolulu, HI 96822, USA}
\affiliation{Subaru Telescope, NAOJ, 650 North A{'o}hoku Place, Hilo, HI 96720, USA}

\author[0000-0001-6305-7272]{Jeffrey Chilcote}
\affiliation{Department of Physics, University of Notre Dame, 225 Nieuwland Science Hall, Notre Dame, IN, 46556, USA}

\author[0000-0003-0156-3019]{Tara Cotten}
\affiliation{Department of Physics and Astronomy, University of Georgia, Athens, GA 30602, USA}

\author{Rene Doyon}
\affiliation{Institut de Recherche sur les Exoplan{\`e}tes, D{\'e}partement de Physique, Universit{\'e} de Montr{\'e}al, Montr{\'e}al QC, H3C 3J7, Canada}

\author[0000-0002-5092-6464]{Gaspard Duch\^ene}
\affiliation{Department of Astronomy, University of California, Berkeley, CA 94720, USA}
\affiliation{Univ. Grenoble Alpes/CNRS, IPAG, F-38000 Grenoble, France}

\author[0000-0002-0176-8973]{Michael P. Fitzgerald}
\affiliation{Department of Physics \& Astronomy, University of California, Los Angeles, CA 90095, USA}

\author[0000-0002-7821-0695]{Katherine B. Follette}
\affiliation{Physics and Astronomy Department, Amherst College, 21 Merrill Science Drive, Amherst, MA 01002, USA}

\author[0000-0003-3978-9195]{Benjamin L. Gerard}
\affiliation{University of Victoria, 3800 Finnerty Rd, Victoria, BC, V8P 5C2, Canada}
\affiliation{National Research Council of Canada Herzberg, 5071 West Saanich Rd, Victoria, BC, V9E 2E7, Canada}

\author[0000-0002-4144-5116]{Stephen J. Goodsell}
\affiliation{Gemini Observatory, 670 N. A'ohoku Place, Hilo, HI 96720, USA}

\author[0000-0002-7162-8036]{Alexandra Z. Greenbaum}
\affiliation{Department of Astronomy, University of Michigan, Ann Arbor, MI 48109, USA}

\author[0000-0003-3726-5494]{Pascale Hibon}
\affiliation{European Southern Observatory, Alonso de Cordova 3107, Vitacura, Santiago, Chile}

\author[0000-0001-9994-2142]{Justin Hom}
\affiliation{School of Earth and Space Exploration, Arizona State University, PO Box 871404, Tempe, AZ 85287, USA}

\author[0000-0003-1498-6088]{Li-Wei Hung}
\affiliation{Natural Sounds and Night Skies Division, National Park Service, Fort Collins, CO 80525, USA}

\author{Patrick Ingraham}
\affiliation{Large Synoptic Survey Telescope, 950N Cherry Ave., Tucson, AZ 85719, USA}

\author[0000-0002-9936-6285]{Quinn Konopacky}
\affiliation{Center for Astrophysics and Space Science, University of California San Diego, La Jolla, CA 92093, USA}

\author[0000-0001-7687-3965]{James E. Larkin}
\affiliation{Department of Physics \& Astronomy, University of California, Los Angeles, CA 90095, USA}

\author{J\'er\^ome Maire}
\affiliation{Center for Astrophysics and Space Science, University of California San Diego, La Jolla, CA 92093, USA}

\author[0000-0001-7016-7277]{Franck Marchis}
\affiliation{SETI Institute, Carl Sagan Center, 189 Bernardo Ave.,  Mountain View CA 94043, USA}

\author[0000-0002-5251-2943]{Mark S. Marley}
\affiliation{NASA Ames Research Center, MS 245-3, Mountain View, CA 94035, USA}

\author[0000-0002-4164-4182]{Christian Marois}
\affiliation{National Research Council of Canada Herzberg, 5071 West Saanich Rd, Victoria, BC, V9E 2E7, Canada}
\affiliation{University of Victoria, 3800 Finnerty Rd, Victoria, BC, V8P 5C2, Canada}

\author[0000-0003-3050-8203]{Stanimir Metchev}
\affiliation{Department of Physics and Astronomy, Centre for Planetary Science and Exploration, The University of Western Ontario, London, ON N6A 3K7, Canada}
\affiliation{Department of Physics and Astronomy, Stony Brook University, Stony Brook, NY 11794-3800, USA}

\author[0000-0001-6205-9233]{Maxwell A. Millar-Blanchaer}
\altaffiliation{NASA Hubble Fellow}
\affiliation{Jet Propulsion Laboratory, California Institute of Technology, Pasadena, CA 91109, USA}

\author[0000-0001-6975-9056]{Eric L. Nielsen}
\affiliation{Kavli Institute for Particle Astrophysics and Cosmology, Stanford University, Stanford, CA 94305, USA}

\author[0000-0001-7130-7681]{Rebecca Oppenheimer}
\affiliation{Department of Astrophysics, American Museum of Natural History, New York, NY 10024, USA}

\author{David Palmer}
\affiliation{Lawrence Livermore National Laboratory, Livermore, CA 94551, USA}

\author{Jennifer Patience}
\affiliation{School of Earth and Space Exploration, Arizona State University, PO Box 871404, Tempe, AZ 85287, USA}

\author[0000-0002-3191-8151]{Marshall Perrin}
\affiliation{Space Telescope Science Institute, Baltimore, MD 21218, USA}

\author{Lisa Poyneer}
\affiliation{Lawrence Livermore National Laboratory, Livermore, CA 94551, USA}

\author{Laurent Pueyo}
\affiliation{Space Telescope Science Institute, Baltimore, MD 21218, USA}

\author[0000-0002-9246-5467]{Abhijith Rajan}
\affiliation{Space Telescope Science Institute, Baltimore, MD 21218, USA}

\author[0000-0003-0029-0258]{Julien Rameau}
\affiliation{Univ. Grenoble Alpes/CNRS, IPAG, F-38000 Grenoble, France}
\affiliation{Institut de Recherche sur les Exoplan{\`e}tes, D{\'e}partement de Physique, Universit{\'e} de Montr{\'e}al, Montr{\'e}al QC, H3C 3J7, Canada}

\author[0000-0002-9667-2244]{Fredrik T. Rantakyr\"o}
\affiliation{Gemini Observatory, Casilla 603, La Serena, Chile}

\author[0000-0003-1698-9696]{Bin Ren}
\affiliation{Department of Physics and Astronomy, Johns Hopkins University, Baltimore, MD 21218, USA}

\author[0000-0003-2233-4821]{Jean-Baptiste Ruffio}
\affiliation{Kavli Institute for Particle Astrophysics and Cosmology, Stanford University, Stanford, CA 94305, USA}

\author[0000-0002-8711-7206]{Dmitry Savransky}
\affiliation{Sibley School of Mechanical and Aerospace Engineering, Cornell University, Ithaca, NY 14853, USA}

\author{Adam C. Schneider}
\affiliation{School of Earth and Space Exploration, Arizona State University, PO Box 871404, Tempe, AZ 85287, USA}

\author[0000-0003-1251-4124]{Anand Sivaramakrishnan}
\affiliation{Space Telescope Science Institute, Baltimore, MD 21218, USA}

\author[0000-0002-5815-7372]{Inseok Song}
\affiliation{Department of Physics and Astronomy, University of Georgia, Athens, GA 30602, USA}

\author[0000-0003-2753-2819]{Remi Soummer}
\affiliation{Space Telescope Science Institute, Baltimore, MD 21218, USA}

\author[0000-0002-5917-6524]{Melisa Tallis}
\affiliation{Kavli Institute for Particle Astrophysics and Cosmology, Stanford University, Stanford, CA 94305, USA}

\author{Sandrine Thomas}
\affiliation{Large Synoptic Survey Telescope, 950N Cherry Ave., Tucson, AZ 85719, USA}

\author[0000-0001-5299-6899]{J. Kent Wallace}
\affiliation{Jet Propulsion Laboratory, California Institute of Technology, Pasadena, CA 91109, USA}

\author[0000-0002-4479-8291]{Kimberly Ward-Duong}
\affiliation{Physics and Astronomy Department, Amherst College, 21 Merrill Science Drive, Amherst, MA 01002, USA}

\author{Sloane Wiktorowicz}
\affiliation{Department of Astronomy, UC Santa Cruz, 1156 High St., Santa Cruz, CA 95064, USA }

\author[0000-0002-9977-8255]{Schuyler Wolff}
\affiliation{Leiden Observatory, Leiden University, 2300 RA Leiden, The Netherlands}

\begin{abstract}

We present a study of the HD 165054 astrometric calibration field that has been periodically observed with the Gemini Planet Imager. HD 165054 is a bright star within Baade's Window, a region of the galactic plane with relatively low extinction from interstellar dust. HD 165054 was selected as a calibrator target due to the high number density of stars within this region ($\sim 3$ stars per square arcsecond with $H<22$), necessary because of the small field-of-view of the Gemini Planet Imager. Using nine epochs spanning over five years, we have fit a standard five-parameter astrometric model to the astrometry of seven background stars within close proximity to HD 165054 ($\rho < 2\arcsec$). We achieved a proper motion precision of $\sim0.3$\,mas\,yr$^{-1}$, and constrained the parallax of each star to be $\lesssim 1$\,mas. Our measured proper motions and parallax limits are consistent with the background stars being a part of the galactic bulge. Using these measurements we find no evidence of any systematic trend of either the plate scale or the north angle offset of GPI between 2014 and 2019. We compared our model describing the motions of the seven background stars to observations of the same field in 2014 and 2018 obtained with Keck/NIRC2, an instrument with an excellent astrometric calibration. We find that predicted position of the background sources is consistent with that measured by NIRC2, within the uncertainties of the calibration of the two instruments. In the future, we will use this field as a standard astrometric calibrator for the upgrade of GPI and  potentially for other high-contrast imagers.  

\end{abstract}

%\keywords{}

\section{Introduction}
\label{sec:intro}

Having an accurate and stable astrometric calibration is necessary for performing precision astrometry with high-contrast imaging. Many instruments take advantage of the high stellar density of globular clusters to perform their astrometric calibration (e.g., \citealp{2016PASP..128i5004S}), but this is not possible with the Gemini Planet Imager (GPI; \citealp{Macintosh:2014js}) for two main reasons: (1) even the brightest stars within these clusters are too faint to act as natural guide stars for the instrument's adaptive optics (AO) system which can only lock onto stars with $I < 10$~mag; (2) GPI has a limited field of view ($2\farcs8\times2\farcs8$), limiting the usefulness of clusters such as the Trapezium \citep{McCaughrean:1994id}.  To address this deficiency, we have observed HD 165054, a star in Baade's Window, using GPI and the Near Infrared Camera 2 (NIRC2) on Keck II. HD 165054 is a good target for use as an astrometric calibrator because it has a high density of visible background sources which can be used to study the astrometric stability of GPI with repeated observations. We also observed this field with NIRC2 to provide an external check of GPI's astrometric solution.

GPI was initially calibrated during commissioning between late-2013 and early-2014 using binary stars such as $\theta^{1}$~Orionis~B in the Trapezium cluster \citep{2014SPIE.9147E..84K}. This study seeks to expand that work by adding HD 165054 to the list of fields that can be used for astrometric calibration. By characterizing this field, we aim to achieve three main science objectives: (1) looking for signs of time-dependent trends in the astrometric calibration of GPI, (2) looking for discrepancies in astrometric calibration between GPI and other high-contrast direct imagers (e.g. VLT/SPHERE, Keck/NIRC2), (3) characterizing the field for continued monitoring of the astrometric calibration of GPI.  GPI is slated to be taken off Gemini South in 2020 and moved to Gemini North following an upgrade of the instrument (colloquially referred to as GPI 2.0). When the instrument is remounted, it will need to be re-calibrated and HD 165054 will be an optimal target for comparing the astrometric calibration before and after the instrument upgrade.

Named after Walter Baade who was the first astronomer to publish images of the field of view in the 1940s, Baade's Window has served as a ``window" into the galactic center due to the field's relatively low concentrations of dust \citep{Baade:1946}. The high density of distant galactic bulge stars visible in Baade's Window makes the field a very good candidate for doing astrometric calibration for GPI. HD 165054, the particular star in Baade's Window that we targeted for this study is a nearby $V=8.48$\,mag G3/5V \citep{1982mcts.book.....H} star that's bright enough for GPI's AO system to lock onto and has a relatively high density of visible "background" stars within the field of view. HD 165054 already has measured absolute astrometry from \textit{Gaia} \citep{GaiaCollaboration:2018io}, but the star is too bright for \textit{Gaia} to discern any other sources in the small $2\farcs8\times2\farcs8$ field around it in which GPI can observe. Using a combination of angular and spectral differential imaging (ADI/SDI \citealp{1999PASP..111..587R,Marois:2006df}), GPI can subtract the point spread function (PSF) of the bright foreground star, allowing us to measure the locations of the faint, distant background stars. 

The paper has been organized into the following sections. In Section \ref{sec:observations}, we give a quick overview of the observations from both GPI and NIRC2. In Section \ref{sec:klip}, we describe the data reduction process for both instruments. In Section \ref{sec:astrometry}, we fit the astrometry of the background stars. This section is divided into two main parts: measuring the positions and fitting for the corresponding proper motions and parallaxes. In Section \ref{sec:discussion}, we analyze the results of the astrometry and discuss them in a broader context. Finally, we conclude in Section \ref{sec:conclusion} with some discussion of future analyses for HD 165054 astrometry and potential new scientific objectives for this field of view. 

\begin{figure*}
    \centering
\minipage{0.50\textwidth}
  \includegraphics[width=\linewidth]{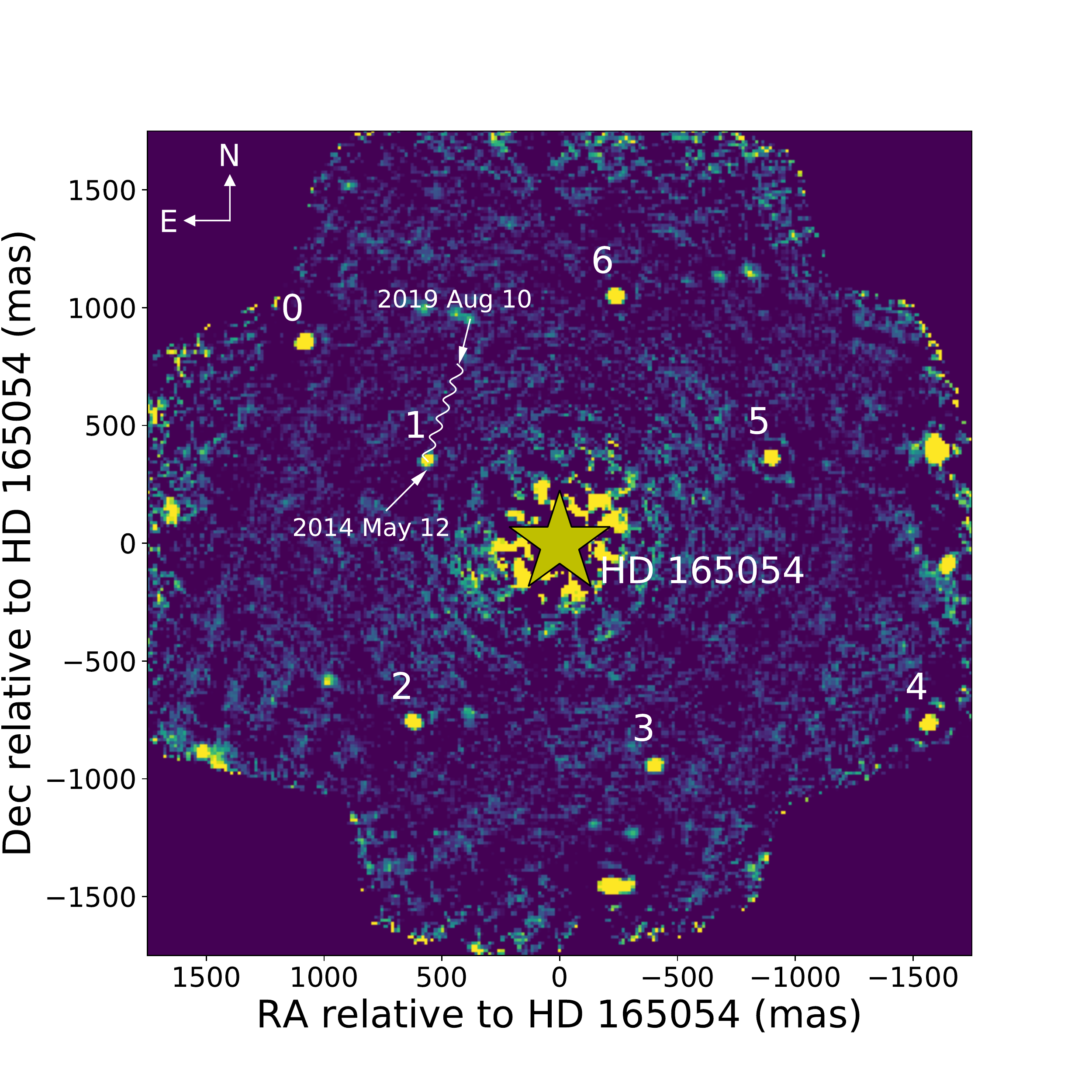}
\endminipage\hfill
\minipage{0.50\textwidth}
  \includegraphics[width=\linewidth]{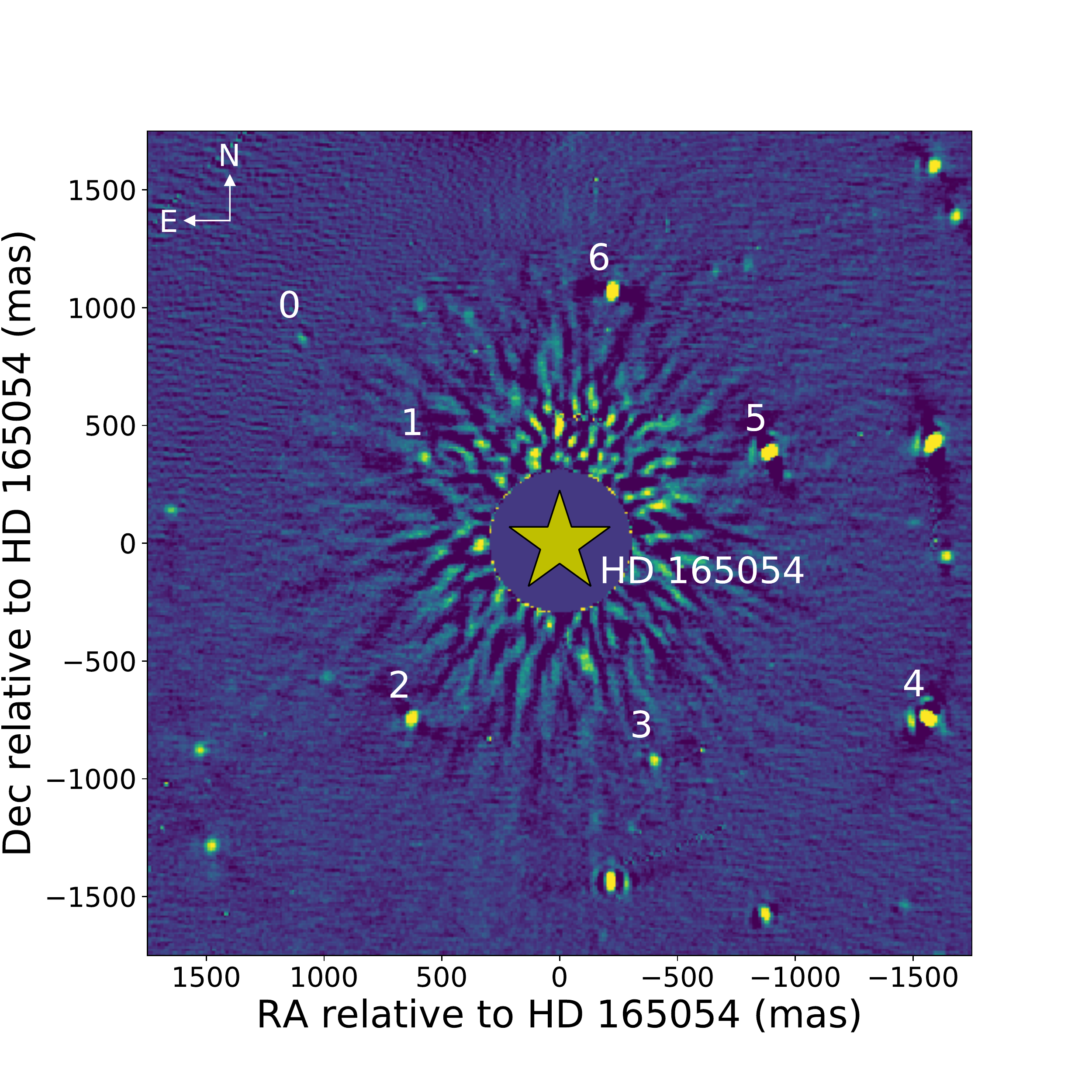}
\endminipage
\caption{HD 165054, observed with GPI on 2014 May 15 (\textbf{Left}) and Keck/NIRC2 on 2014 Jul 12 (\textbf{Right}). The seven brightest background stars around HD 165054 that we characterized are given simple ordered numeric labels. The best-fit model of the position of background star 1 over time is plotted as a white track to help visualize the apparent relative motion of the field over the approximately five year baseline of observations (2014 May 12 - 2019 Aug 10).
\label{fig:pyklip}}
\end{figure*}

\section{Observations and Data Reduction}
\label{sec:observations}

\subsection{Gemini South/GPI}
The Gemini Planet Imager (GPI, \citealp{Macintosh:2014js}) is a high-contrast instrument designed to detect faint substellar companions at small angular separations from their bright host stars. The instrument combines a high-order adaptive optics system to achieve near-diffraction limited imaging \citep{2014SPIE.9148E..0KP}, an apodized Lyot coronagraph to suppress the light from the central star \citep{2010SPIE.7735E..86S,2011ApJ...729..144S}, and an integral field spectrograph \citep{Chilcote:2012hd,2014SPIE.9147E..1KL} to obtain a low-resolution ($R\approx 45$) spectrum at each point within the field of view. HD 165054 has been observed nine times during the course of the Gemini Planet Imager Exoplanet Survey (GPIES; \citealp{Nielsen:2019cb}), providing us with a temporal baseline of approximately five years (see Fig. \ref{fig:pyklip} for a sample derotated and collapsed image of HD 165054 using GPI). Each dataset was obtained in the standard $H$-band coronagraphic mode, although the number of frames varied between epochs. An observing log is given in Table~\ref{tbl:log} listing the number of images obtained, the achieved field rotation, and the average environmental conditions. 

\startlongtable
\begin{deluxetable*}{cccccccccc}
\tablecaption{Observation Log\label{tbl:log}}
\tablehead{
\colhead{Date of} & \colhead{UTC at} &  \colhead{Instrument} & \colhead{Filter} & \colhead{Number} & \colhead{Frame} & \colhead{Field} & \colhead{True} & \colhead{Average} & \colhead{Average}\\
observation & start of & & & of & exposure & rotation & north & DIMM & MASS\\
& observation & & & frames & time & & offset & seeing & seeing\\
& & & & & (s) & (deg) & (deg) & (arcsec) & (arcsec)}
\startdata
2014 May 12 & 07:50:29.7 & GPI & $H$ & 10 & 60 & 5.6 & $0.23\pm0.11$ & 1.4 & 0.76\\ 
2014 May 15 & 06:39:46.7 & GPI & $H$ & 37 & 60 & 149.7 & $0.23\pm0.11$ & 0.71 & 0.64\\
2014 Jul 12 & 08:28:15.7 & NIRC2 & $K^{\prime}$ & 87 & 30 & 18.7 & $-0.252\pm0.009$ & \nodata$^{\tablenotemark{a}}$ & \nodata$^{\tablenotemark{a}}$\\
2015 Jul 3 & 04:17:27.8 & GPI & $H$ & 18 & 60 & 46.7 & $0.17\pm0.14$ & 1.22 & 0.54\\
2015 Sep 1 & 00:42:25.4 & GPI & $H$ & 33 & 60 & 5.3 & $0.17\pm0.14$ & 1.31 & 1.31\\
2016 Apr 30 & 08:28:53.7 & GPI & $H$ & 38 & 60 & 15.9 & $0.21\pm0.23$ & 2.24 & 1.38\\
2017 Aug 7 & 01:31:25.3 & GPI & $H$ & 25 & 60 & 136.6 & $0.32\pm0.15$ & \nodata & \nodata \\
2018 Jul 21 & 09:11:41.1 & NIRC2 & $K^{\prime}$ & 63 & 30 & 12.6 & $-0.262\pm0.020$ & \nodata & \nodata\\
2018 Aug 10 & 01:03:49.0 & GPI & $H$ & 34 & 60 & 148.5 & $0.28\pm0.19$ & \nodata & \nodata \\
2019 Mar 29 & 10:07:54.2 & GPI & $H$ & 11 & 60 & 124.2 & $0.45\pm0.11$ & \nodata & \nodata \\
2019 Aug 10 & 01:02:20.9 & GPI & $H$ & 38 & 60 & 117.3 & $0.45\pm0.11$ & \nodata & \nodata \\
\enddata
\tablenotetext{a}{Environmental data not available}
\end{deluxetable*}

All raw 2-D data files were reduced with the GPI Data Reduction Pipeline (DRP v1.5.0; \citealp{Perrin:2014jh, Perrin:2016}) using the same recipe described as follows. First, a dark background was subtracted. Then bad pixels were interpolated over. Instrument flexure was corrected for using an argon arc lamp calibration \citep{Wolff:2014cn}. A 3-D datacube was assembled along the wavelength axis using 2-D image spectra. This newly created wavelength axis was then interpolated over to produce 37 evenly spaced wavelength channels. After this, a second bad pixel interpolation was applied to the 3-D datacube. Then, a geometric distortion correction was applied \citep{2014SPIE.9147E..84K}. Finally, the locations and flux of satellite spots were measured. The satellite spots are four copies of the stellar PSF of the star behind the coronagraph. These satellite spots are attenuated in flux and distributed in a radially symmetric pattern around the target star. The positions of the satellite spots are measured in order to triangulate the position of the target star since the coronagraph blocks our ability to measure the star's position directly. We used a plate scale of $14.161\pm0.021$\,mas\,px$^{-1}$ for each epoch, and a variable north offset angle given for each epoch in Table~\ref{tbl:log}, defined as $\theta_{\rm true} - \theta_{\rm observed}$ \citep{2019arXiv191008659D}.

\subsubsection{Parallactic Angle Correction}
\label{sec:parallactic angle correction}
Imperfections in the construction of the Gemini South telescope cause an offset between the true vertical axis and the $y$-axis of the telescope, an effect that is most pronounced when observing a target at a very small zenith distance. This causes an offset in the measured and actual parallactic angle, biasing position angle measurements made from a GPI image. While this offset is usually corrected for with the Cassegrain derotator upon which GPI is mounted, there have been several GPI observing runs where this rotator drive is disabled, and it remains fixed in one position.

\begin{figure}
\includegraphics[width=\columnwidth]{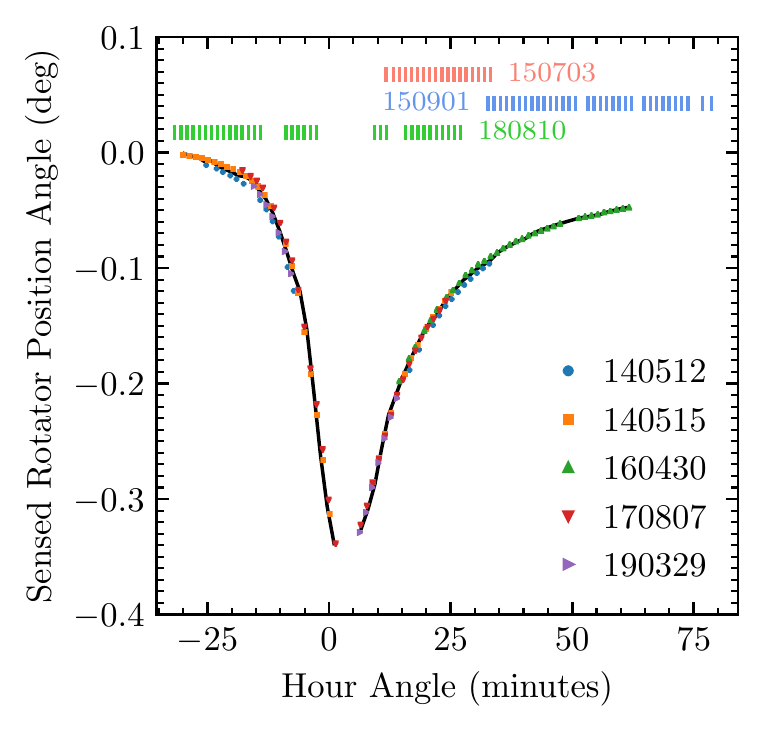}
\caption{Sensed position of Gemini South's Cassegrain rotator as a function of hour angle recorded in the header for observations of HD 165054 obtained with the rotator drive enabled. The symbols denote the measurements for the different epochs (see inset legend). A model of the rotator position angle as a function of hour angle (black line) was constructed from these measurements. The model is incomplete near zenith due to the lack of measurements. The hour angles of the frames in the three epochs where the rotator drive was disabled are indicated at the top of the plot.\label{fig:crpa}}
\end{figure}
This effect is significant for HD 165054. At a declination of $-28^{\circ}41^{\prime}10.8^{\prime\prime}$, the star has a minimum zenith distance of $1\fdg6$ as it transits the observatory. Fig.~\ref{fig:crpa} shows the change in the sensed angle of the Cassegrain rotator as a function of hour angle for the observations of this star where the rotator drive was enabled. The correction required to keep vertical aligned with the $y$-axis in the image plane ranges between $0\degr$ and $-0\fdg35$, depending on the current hour angle. 

For three of the epochs we obtained on HD 165054 (2015 July 03, 2015 Sep 01 and 2018 Aug 10) the Cassegrain rotator drive was disabled causing the vertical angle to drift during the course of the observing sequence. As our observations of HD 165054 are timed to observe the star near transit, this offset can lead to a significant bias in the measured position angle of the background stars relative to the foreground star. To correct these three datasets we constructed a model of the required rotator position angle to correct the vertical angle from the five epochs where the rotator drive was enabled (Fig. ~\ref{fig:crpa}). This model was constructed by taking the median sensed angle over a small hour angle range.

We corrected the parallactic angle for each image obtained during these three epochs by subtracting the predicted Cassegrain rotator position angle calculated using this model from the average parallactic angle calculated by the DRP. For the subset of the frames in the 2015 Sep 01 epoch where the hour angle was outside of the range of the model, we performed a linear extrapolation to predict the correction to apply to the parallactic angle. In De Rosa et al. 2019 (\textit{submitted}) we used a different approach to estimate the correction to be applied to the parallactic angle based on a simple model of the telescope non-perpendicularity. The difference between these two approaches was very small for the observations of HD 165054; the median absolute difference was 0.005\,deg, and 79 of the 85 frames had a difference smaller than 0.02\,deg.

\subsection{Keck II/NIRC2}

Two epochs of HD 165054 imaging data were obtained (see Table \ref{tbl:log}) using Keck/NIRC2 and the facility's AO system. We observed HD 165054 using NIRC2 to serve as an external check of GPI's astrometry because NIRC2 has a well-calibrated astrometric solution tied to observations of globular clusters that have been observed with HST \citep{Yelda:2010ig,2016PASP..128i5004S}. The raw data were reduced using a standard near-infrared data reduction pipeline. First, a dark background was subtracted from the science frames, and then a flat field correction was applied. After this, bad pixels in the science frame were interpolated over with cubic spline interpolation. These bad pixels were identified by looking for pixels in the flat field frame that were 5$\sigma$ discrepant from adjacent pixels within a 20-pixel search box. Finally, a geometric distortion correction was applied using the appropriate NIRC2 narrow camera distortion solution \citep{Yelda:2010ig, 2016PASP..128i5004S}.

After reducing the flat-fielded frames, we measured the position of the foreground star behind the coronagraphic mask. While the occulting mask is semi-transparent, the position of the attenuated spot may not accurately reflect the true location of the star within the image plane (typical systematic offsets of 0.2\,px have been measured). Instead, we used a two-step iterative method that determines the relative alignment of the images within each dataset and the absolute location of the star after stacking the aligned images.

In the first step, the relative alignment of the foreground star between each frame was determined by fitting the path of the background stars through the sequence of images. The background stars should each follow the path of a circle centered at the location of the foreground star, as the rotation axis of NIRC2 is centered on the guide star. We constructed a model with $2n_{\rm bkg} + 2$ parameters; a radius and position angle in the first image of a circle describing the path of each background object, and the ($x$, $y$) coordinate of the foreground star, defining the center of each circle. There were 13 and 18 background objects used in this analysis for the 2014 and 2018 datasets, respectively. We used the location of the spot within the coronagraphic mask as an initial guess of the star position within each image. After finding the best fit combination of radii, position angles, and star center position, we modified the initial guess of the star position within each image by the average residual of the path of the background stars relative to the best fit model. This process was then repeated using the revised guess for the star center location within each image.

The images within each dataset were now aligned relatively to one another, but the absolute star center was not well constrained due to the limited amount of field rotation achieved for each dataset. The second step of the alignment procedure was measuring the location of the star position within a median-combined image of the full dataset. Taking the median of the stack without derotating the images to put North up results in a high signal-to-noise-ratio (SNR) image of the diffraction pattern of the telescope; background stars are removed as they only occupy a given pixel for a small fraction of the full observing sequence. We applied the radon transform method described in \citet{Pueyo:2015cx} on this median-combined image to measure the absolute star center (the radon transform assumes the diffraction spikes point towards the true center of the star and measures flux radially along these spikes to find this position). We then repeated the two-step process, but this time fixing the two free parameters describing the star center to the values measured using the radon transform. This iterative procedure was repeated three times, after which the location of the star center, and the relative alignment between the images, no longer changed.

\section{PSF Subtraction and Point Source Astrometry}
\label{sec:klip}
After reducing the raw science frames and transforming them into datacubes using the GPI DRP, we then subtracted the foreground star PSF using the \texttt{pyKLIP} \citep{Wang:2015pyklip} Python implementation of Karheunen-Loeve Image Processing (KLIP) \citep{2012ApJ...755L..28S,Pueyo:2015cx} (see Fig. \ref{fig:pyklip}), using seven Karheunen-Loeve (KL) modes for the PSF subtraction. The reference library used to do the PSF subtraction was chosen such that ADI/SDI will have caused any astrophysical sources to move by 4 pixels.  We used the same parameters to do PSF subtraction for both the GPI and NIRC2 data. Then, the PSF-subtracted images at each wavelength were derotated and combined into a single image using a weighted-mean.

Since none of the background stars being studied have been characterized or catalogued yet, we have given them placeholder numerical labels from 0 to 6 in order of increasing position angle. The background stars have contrasts relative to HD 165054 of $10.0\le\Delta H\le12.5$. The positions were fitted for using an algorithm called Bayesian-KLIP Astrometry (BKA, \citealt{Wang:2015pyklip}). This method takes the PSF-subtracted images generated using KLIP and then employs a Bayesian framework using a Markov Chain Monte Carlo (MCMC) sampler along with forward modeling of the background star's PSF in order to fit for that star's position.

The first step in BKA involves generating a template for forward modeling called an \textit{instrumental PSF}. For the GPI data, this instrumental PSF is generated by averaging together the satellite spots produced in the data reduction step (see Section 2). For the NIRC2 data, this instrumental PSF was extracted from a bright background source for the first epoch (not one of the seven background stars being analyzed for astrometry), and from the unocculted foreground star in the second epoch (unfortunately the foreground star PSF could not be extracted in the first epoch). Generating the instrumental PSF for a given epoch of NIRC2 data involves extracting a 17-by-17 pixel ``postage stamp'' from each image and then averaging these stamps together to maximize SNR.

Once an instrumental PSF is created, the next step is to measure the positions of the seven background stars being characterized. For each background star, we extracted an 11-by-11 pixel data stamp of the star based on an initial guess position and fit a forward modeled PSF (generated using the instrumental PSF) to it. We fit the forward model to the data using a Gaussian likelihood function to generate a posterior distribution of the background star's position.  See Fig. \ref{fig:bka_example} for an example fit (see Appendix \ref{sec:measured positions} for a table of all measured background star positions for each epoch).

The background and foreground star positions are measured using different techniques (the background star positions using this BKA algorithm, the foreground star position using the satellite spots), but the relative astrometry we obtain properly accounts for and constrains any systematics that could be due to using these different techniques. Our astrometric error budget properly adds the uncertainties on these two techniques in quadrature, and the residuals we obtain at the end of our analysis (See Fig. \ref{fig:residuals}) would constrain any systematics down to approximately the weighted-mean of the error bars on the residuals ($\sim 3$ mas). For example, if the satellite spots for our foreground star centering were all systematically offset in one direction, we would see a similar systematic offset in our residuals.

\begin{figure*}
    \centering
    \includegraphics[width=1.0\textwidth]{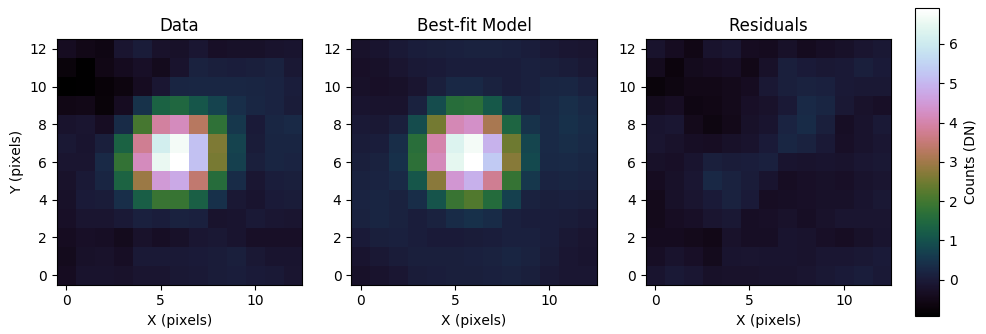}
    \caption{Sample output of data, best-fit model, and residuals of Bayesian KLIP Astrometry (BKA) for the background star 0 during the first epoch (2014 May 5). Given an initial guess of star position, a 13-by-13 pixel stamp is extracted from the science f
    rame (\textbf{Left}). Then, a best-fit model (\textbf{Middle}) based on forward modeling of the stellar PSF is used in a Gaussian likelihood function to generate a posterior distribution of the star's position. Residuals (\textbf{Right}) shown for comparison.\label{fig:bka_example}}
\end{figure*}

\section{Fitting for Proper Motion and Parallax}
\label{sec:astrometry}

Once the detector positions of the background stars were obtained in each image, the positions were then measured relative to the foreground star position in order to place the background stars in the foreground star's reference frame. The motion of the foreground star was modeled using \textit{Gaia} DR2 measurements of HD 165054. 

\subsection{Equations of motion}
\label{sec:equations of motion}
The position of a star due to its proper motion and parallax as a function of time is described by the following equations: 

\begin{equation}
\label{eqn:motion}
\begin{split}
    \alpha^{\star}(t) &= \mu_{\alpha^{\star}} t + \pi\left(X\sin\alpha_0 - Y\cos\alpha_0\right)\\
    \delta(t) &= \mu_\delta t + \pi\left(X\cos\alpha_0\sin\delta_0 + Y\sin\alpha_0\sin\delta_0 - Z\cos\delta_0\right)
\end{split}
\end{equation}

where $t$ is the time relative to a reference epoch (chosen to be 2016 Oct 10 as this is the date exactly halfway between the first and last epoch of observations); $\alpha^{\star}$ and $\delta$ are the relative change in right ascension (RA) and declination (Dec) respectively; $\alpha_0$ and $\delta_0$ are the RA and Dec at the reference epoch; $\mu_{\alpha^{\star}}$ and $\mu_\delta$ are the proper motion in RA and Dec respectively; $\pi$ is the parallax; and $X$, $Y$, and $Z$ are the barycentric coordinates of the Earth. Here we use the notation $\alpha^{\star}=\alpha\cos\delta$ to describe coordinates in the tangent plane, and we use milliarcseconds for all angular quantities, years for time, and au for the barycentric position of the Earth. We use the function \texttt{get\_body\_barycentric} from the \texttt{astropy.coordinates} Python package \citep{TheAstropyCollaboration:2013cd} to conveniently calculate $X$, $Y$, and $Z$ for us as a function of time. 

These background stars are measured relative to HD 165054, so we need to simultaneously model the proper and parallactic motions of HD 165054 as well as the motion of the background star in order to obtain absolute astrometry of the background star. Given that these measurements are also relative to a reference epoch, we must account for the position of the star at this reference epoch. The relative position of a background star can be expressed as a function of the absolute astrometry of both foreground and background star as
\begin{equation}
\begin{split}
\alpha^{\star}_{\rm rel.}(t) &= \alpha^{\star}_{\rm bkg}(t) - \alpha^{\star}_{\rm HD165}(t)\\
\delta_{\rm rel.}(t) &= \delta_{\rm bkg}(t) - \delta_{\rm HD165}(t)
\end{split}
\end{equation}
where $\alpha^{\star}_{\rm bkg}(t)$ and $\alpha^{\star}_{\rm HD165}(t)$ are the absolute astrometric motions of the background and foreground star, respectively. This results in ten astrometric parameters that describe the motion of a given background star relative to HD 165054. For the foreground star, the values and associated uncertainties for the five parameters $(\alpha, \delta, \pi,  \mu_{\alpha^{\star}}, \mu_\delta)_{\rm HD165}$ were taken from the \textit{Gaia} DR2 catalog \citep{GaiaCollaboration:2018io}, reported at the reference epoch of 2015.5 (2015 Jul 2). These parameters precisely describe the position and apparent motion of the foreground star across the sky.

The corresponding suite of five parameters describing the background star $(\alpha, \delta, \pi, \mu_{\alpha^{\star}}, \mu_\delta)_{\rm bkg}$ have three unknown parameters that must be fitted for using MCMC. Seeing as we are measuring the background star positions on a flat plane tangent to the RA and Dec of the foreground star, deviations from this flat tangent plane approximation (which are on the order of $\sim 50$ nano-arcseconds) are negligible given the apparent separation of the background stars from the foreground star. We include two additional parameters, an offset in RA ($\Delta\alpha^{\star}$) and Dec ($\Delta\delta$), to properly account for the uncertainty on the true position of the star at the reference epoch.

\subsection{Parameter estimation}
\label{sec:parameter estimation}
We used MCMC to fit these ten astrometric parameters; five describing the astrometry of the background star and five describing the astrometry of the foreground star. We performed this fit using only the GPI data because there were too few epochs of NIRC2 data to constrain these astrometric parameters with that instrument. Although the five parameters describing the astrometry of the foreground star are well-constrained and will be marginalized over as nuisance parameters, we include them here for completeness. We used the affine-invariant sampler from the Python package \texttt{emcee} \citep{ForemanMackey:2013io} to efficiently sample the posterior distributions of each of these parameters. We adopted Gaussian priors for the five parameters describing the astrometry of HD 165054, using the measurements and uncertainties from the {\it Gaia} catalogue. Uniform priors were used for the proper motion and reference epoch offset parameters for the background star. For the parallax of the background star we used a prior from \citet{2015PASP..127..994B}, which we discuss more in Sec.~\ref{sec:parallax}. We initialized 256 walkers randomly throughout parameter space and advanced the chains for 5000 steps, discarding the first 2000 steps as a ``burn-in''. Convergence of the chains was visually assessed by eye to be satisfactory.

\section{Discussion/Results} \label{sec:discussion}

The ten-parameter MCMC fit produced Gaussian posterior distributions for almost all of the  parameters (except for parallax) for each background star. This was expected since the motion we are fitting is simple parallactic and proper motion. See Table \ref{tbl:MCMC fit} for the median and 1-$\sigma$ values for the posterior distributions of each MCMC parameter. A gallery that shows best-fit model curves based on parameters sampled from the posterior distributions is provided in Fig.~\ref{fig:gallery}.

\startlongtable
\begin{deluxetable*}{cc|ccccccc}
\label{tbl:MCMC fit}
\tablecaption{Median and 1-$\sigma$ confidence intervals of the astrometric parameters for each background star}
\tablehead{& & \multicolumn{7}{c}{Background star number}\\
\colhead{Parameter} & \colhead{Units} & \colhead{0} & \colhead{1} & \colhead{2} & \colhead{3} & \colhead{4} & \colhead{5} & \colhead{6}}
\startdata
$\mu_{\alpha^{\star}} $ & $(mas/yr)$ & $-5.44_{-0.28}^{+0.29}$ & $-2.03_{-0.23}^{+0.24}$ & $0.2_{-0.23}^{+0.23}$ & $2.97_{-0.26}^{+0.26}$ & $-3.48_{-0.28}^{+0.27}$ & $-2.24_{-0.20}^{+0.21}$ & $-2.13_{-0.25}^{+0.26}$ \\
$\mu_\delta$ & $(mas/yr)$ & $-5.56_{-0.24}^{+0.24}$ & $-8.94_{-0.21}^{+0.21}$ & $-5.73_{-0.21}^{+0.21}$ & $-3.36_{-0.21}^{+0.21}$ & $-5.18_{-0.24}^{+0.24}$ & $-2.60_{-0.21}^{+0.21}$ & $-5.98_{-0.21}^{+0.21}$ \\
$\pi$ & $(mas)$ & $0.31_{-0.11}^{+0.21}$ & $0.280_{-0.096}^{+0.160}$ & $0.24_{-0.08}^{+0.12}$ & $0.32_{-0.12}^{+0.21}$ & $0.36_{-0.15}^{+0.28}$ & $0.29_{-0.10}^{+0.17}$ & $0.53_{-0.25}^{+0.48}$ \\
$\Delta \alpha^{\star}$ & $(mas)$ & $1014.57_{-0.52}^{+0.52}$ & $501.66_{-0.38}^{+0.37}$ & $568.03_{-0.33}^{+0.33}$ & $-453.77_{-0.41}^{+0.42}$ & $-1628.15_{-0.47}^{+0.46}$ & $-956.3_{-0.30}^{+0.30}$ & $-293.59_{-0.45}^{+0.44}$ \\
$\Delta \delta$ & $(mas)$ & $1053.79_{-0.41}^{+0.41}$ & $547.39_{-0.34}^{+0.34}$ & $-554.76_{-0.34}^{+0.35}$ & $-727.71_{-0.35}^{+0.35}$ & $-555.41_{-0.40}^{+0.40}$ & $579.24_{-0.35}^{+0.35}$ & $1254.39_{-0.35}^{+0.35}$ \\
\hline
$\Delta \alpha_{\rm fgd}$ & $(mas)$ & $0.001_{-0.067}^{+0.068}$ & $0.000_{-0.067}^{+0.067}$ & $0.000_{-0.066}^{+0.067}$ & $-0.001_{-0.066}^{+0.068}$ & $0.002_{-0.068}^{+0.068}$ & $0.000_{-0.067}^{+0.067}$ & $-0.001_{-0.067}^{+0.068}$ \\
$\Delta \delta_{\rm fgd}$ & $(mas)$ & $0.001_{-0.059}^{+0.059}$ & $0.000_{-0.059}^{+0.060}$ & $-0.001_{-0.059}^{+0.059}$ & $0.002_{-0.059}^{+0.059}$ & $0.001_{-0.059}^{+0.058}$ & $-0.001_{-0.059}^{+0.059}$ & $0.000_{-0.059}^{+0.059}$ \\
$\Delta \pi_{\rm fgd}$ & $(mas)$ & $0.005_{-0.062}^{+0.062}$ & $0.010_{-0.061}^{+0.061}$ & $0.021_{-0.062}^{+0.061}$ & $0.003_{-0.061}^{+0.061}$ & $0.001_{-0.061}^{+0.061}$ & $0.008_{-0.061}^{+0.061}$ & $-0.007_{-0.061}^{+0.062}$ \\
$\Delta \mu_{\alpha^{\star}, fgd}$ & $(mas/yr)$ & $0.00_{-0.14}^{+0.14}$ & $0.00_{-0.15}^{+0.15}$ & $0.00_{-0.15}^{+0.14}$ & $0.00_{-0.14}^{+0.15}$ & $0.00_{-0.14}^{+0.15}$ & $0.00_{-0.15}^{+0.14}$ & $0.00_{-0.14}^{+0.14}$ \\
$\Delta \mu_{\delta, {\rm fgd}}$ & $(mas/yr)$ & $0.00_{-0.12}^{+0.12}$ & $0.00_{-0.12}^{+0.12}$ & $0.00_{-0.12}^{+0.12}$ & $0.00_{-0.12}^{+0.12}$ & $0.00_{-0.12}^{+0.12}$ & $0.00_{-0.12}^{+0.12}$ & $0.00_{-0.12}^{+0.12}$ \\
\enddata
\tablecomments{All foreground star astrometric parameters are given as differential measurements (by subtracting off the measured Gaia DR2 astrometry for the foreground star, HD 165054)}
\end{deluxetable*}

\subsection{Parallax}
\label{sec:parallax}

\begin{figure}
    \centering
    \includegraphics{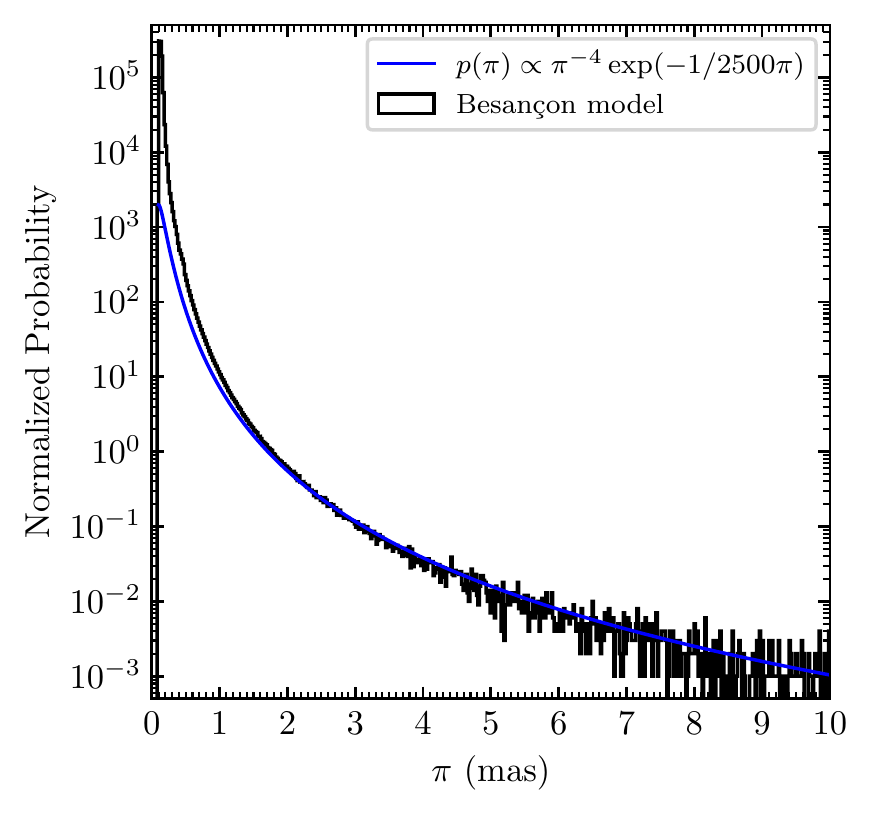}
    \caption{Comparison of the predicted distribution of stars from the BGM (black histogram) to the Bailer-Jones prior (blue), assuming an exponential cut-off distance of $L=2500$\,pc.
    \label{fig:plx_prior}}
\end{figure}

To fit the parallax, we used a physically motivated prior based on galactic population models \citep{2015PASP..127..994B} that combines a constant volume density assumption with a decaying exponential function. If you assume a uniform volume density, the number of stars we would expect to find within an infinitesimally thin spherical shell of radius $r$ scales as $r^2$. Because parallax scales with inverse distance, this means that the number of stars in the corresponding region of parallax parameter space scales as $p(\pi) \propto 1/\pi^4$. This prior does not account for the increase in volume density of stars due to the increasing concentration of stars towards the galactic center (see Fig.~\ref{fig:plx_prior}). 

\begin{figure}
    \centering
    \includegraphics[width=\columnwidth]{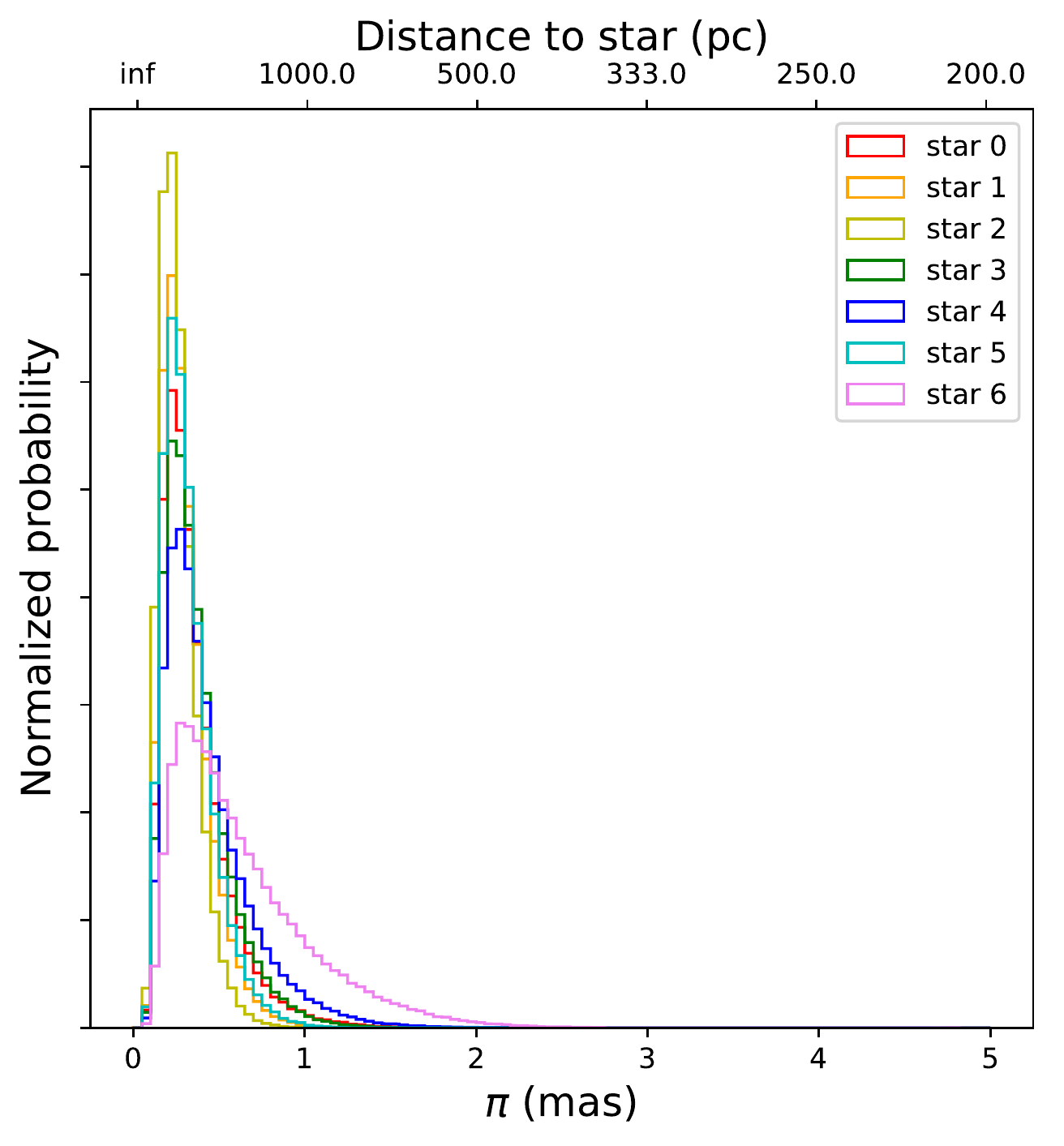}
    \caption{Parallax posterior distributions for the background stars using the prior from \citet{2015PASP..127..994B}.}
\end{figure}

The exponentially decreasing constant volume density prior has the following analytic form (see \citealt{2015PASP..127..994B} for the full derivation):

\large
\begin{equation}
\label{eqn:prior1}
    p(r) = 
    \begin{cases}
        C\frac{1}{2L^3} r^{2} e^{-r/L} & \text{if} \hspace{3mm} r > 0\\
        0 & \text{otherwise}
    \end{cases}
\end{equation}
\normalsize

Where \textit{r} is distance, \textit{L} is a tunable free parameter corresponding to a length scale, and $C$ is a normalization constant. We are most interested in parallax, so Equation~\ref{eqn:prior1} is more useful to us if we convert \textit{r} to \textit{$\pi$}. An algebraic transformation gives us the following equation: 

\large
\begin{equation}
    p(\pi) =
    \begin{cases}
        C\pi^{-4} e^{-1/\pi L} & \text{if} \hspace{3mm} \pi > 0\\
        0 & \text{otherwise}
    \end{cases}
\end{equation}
\normalsize
where $\pi$ is expressed in arcseconds, and $L$ in parsecs. To check the validity of this prior, we used the Besan\c{c}on galactic model (BGM, \citealp{2014A&A...564A.102C}) to simulate the population of stars in the direction of HD 165054. We simulated a region with a solid angle of 0.01 deg$^2$ out to a distance of 10\,kpc. We did not apply any constraints on the apparent or absolute magnitudes or colors of the simulated stars. The output of this simulation was a population of ${\sim}10^6$ stars distributed according to the model of the Galaxy within the Besan\c{c}on model. We compared the predicted distribution of observed stars as a function of parallax between the BGM simulation and the Bailer-Jones prior in Fig.~\ref{fig:plx_prior}. We found $L=2500$\,pc to be a good fit to the simulated distribution. The Bailer-Jones prior using this value matches the distribution produced by the BGM well over the range of parallax values to which we are sensitive (${\gtrsim}1$ mas). Since the Bailer-Jones prior does not take into account the concentration of stars in the central bulge, there is a discrepancy between the BGM and Bailer-Jones model at small parallax values (${\lesssim}1$\,mas). Fortunately, this discrepancy occurs outside of the range of sensitivity of our relative astrometry ($\pi > 1$\,mas), so the prior remain valid for our purpose in this study. 

By comparing the parallax posterior distributions produced by a uniform vs. the Bailer-Jones prior, it was apparent that the Bailer-Jones prior dominated the behaviour of the MCMC fit, indicating that the limiting factor on constraining the parallaxes is the fact that GPI is not sensitive to parallaxes below a $\sim 1$\,mas. The parallax posterior distribution implies that the background stars are most likely all central bulge stars and that our models do not have much more ability to constrain the parallax/distance beyond that. Fortunately, we do not need to constrain the parallax beyond this point to get good astrometry of the background stars. Our MCMC fit constrained the proper motions of the background stars equally well with either prior (uniform or Bailer-Jones), and as will be discussed in Sec. \ref{sec:proper motion}, the implication that these background sources are all central bulge stars is consistent with the proper motions that we measure. The Bailer-Jones prior was assumed to be independent of flux because, as we can see in Fig. \ref{fig:plx_vs_V}, the parallaxes of the stars simulated using the Besan\c{c}on model are well below GPI's $\sim 1$ mas parallax detection limit.

\subsection{Proper Motion}
\label{sec:proper motion}

\begin{figure}
    \centering
    \includegraphics[width=\linewidth]{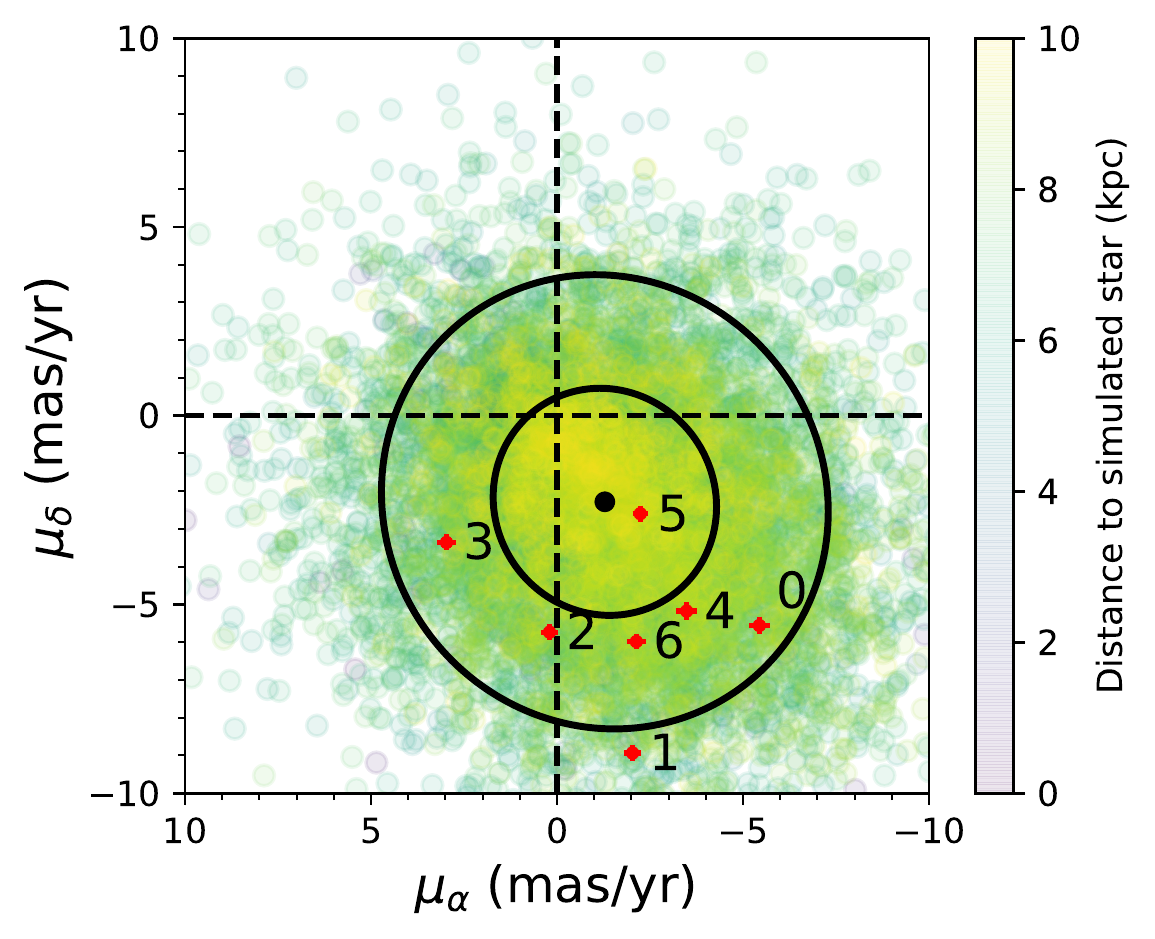}
    \caption{Proper motions of characterized background stars (red points) plotted over a simulated distribution of proper motions for a sample of ${\sim}10^5$ stars in a 0.01 deg$^2$ field of view around HD~165054.\label{fig:proper motions}}
\end{figure}

Before performing the astrometry, the naive assumption was that the background stars being studied were probably all central bulge stars with proper motions consistent with galactic rotation around the center of the Milky Way. This assumption would help explain the similar direction and magnitude of proper motion that we measured for the background stars (see Fig.~\ref{fig:proper motions}). The BGM model that we used to simulate the distribution of galactic star parallaxes was also used to simulate the distribution of proper motions of stars in the direction of HD 165054. We find that the proper motions of the background stars are consistent with stars approximately 5-10 kpc away (Fig. \ref{fig:proper motions})--- the approximate distance from our solar system to the galactic center. Though this is not a rigorous constraint on the actual distance of the background stars, it is a good check to show that the proper motions we measure are consistent with our estimates for their parallax.

\subsection{Residuals}
\label{sec:residuals}
\begin{figure*}
    \centering
    \includegraphics[width=\textwidth]{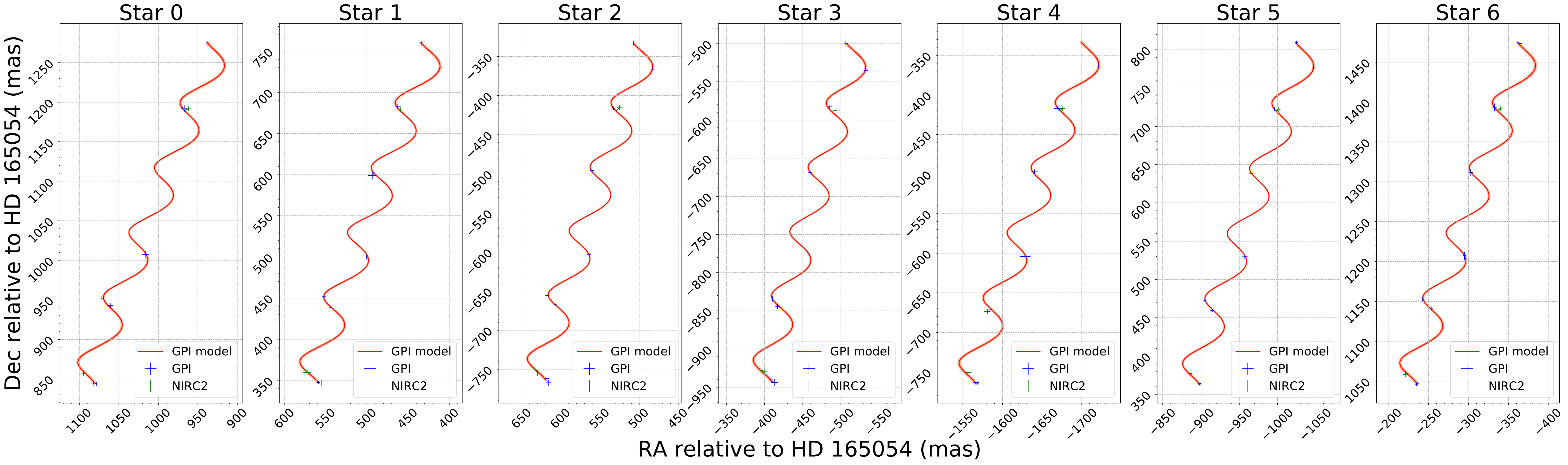}
\caption{Gallery of best-fit model tracks of background stars. The bold red track is the mean of the 100 fainter tracks plotted. Each of the tracks is generated with a random sample from the posterior distributions of the MCMC parameters. A line is drawn for each data point connecting the measured position of the star to the mean predicted position of the star at that epoch.\label{fig:gallery}}
\end{figure*}

As a first test of the validity of our results, we took a look at the residuals of the astrometric fits. To produce the residuals, we first generated best-fit model curves that traced out the predicted positions of the background stars over time (see Fig. \ref{fig:gallery}). To produce these curves, we took our model of the parallactic and proper motion of the background stars relative to the reference epoch and then added on the fitted position of the background star at the reference epoch (the RA ($\Delta\alpha^{\star}$) and Dec ($\Delta\delta$) offset terms used in the MCMC fit, see Sec. \ref{sec:equations of motion}). 

\begin{figure*}
    \centering
\minipage{0.5\textwidth}
  \includegraphics[width=\linewidth]{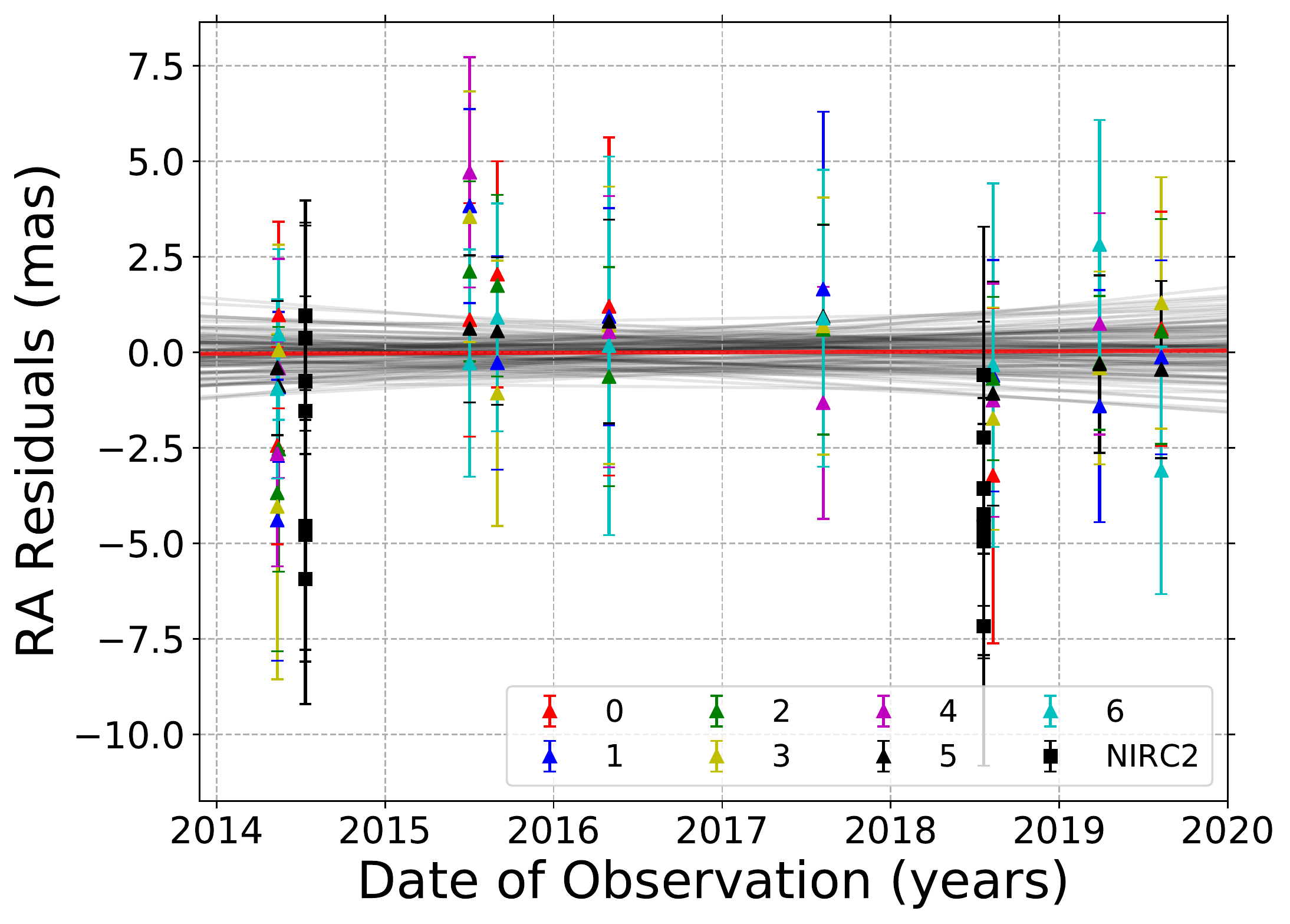}
\endminipage\hfill
\minipage{0.5\textwidth}
  \includegraphics[width=\linewidth]{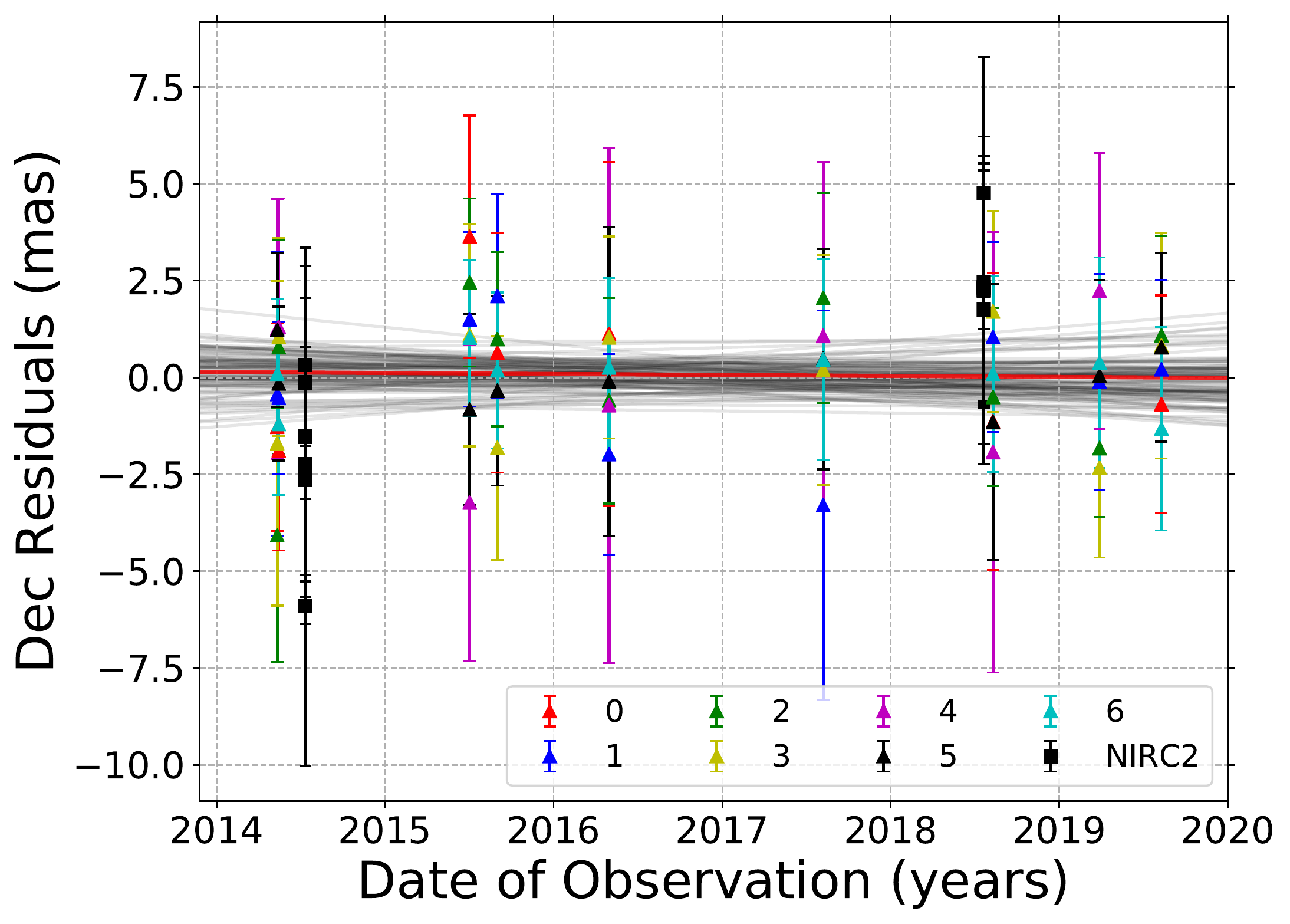}
\endminipage\hfill
\minipage{0.5\textwidth}
  \includegraphics[width=\linewidth]{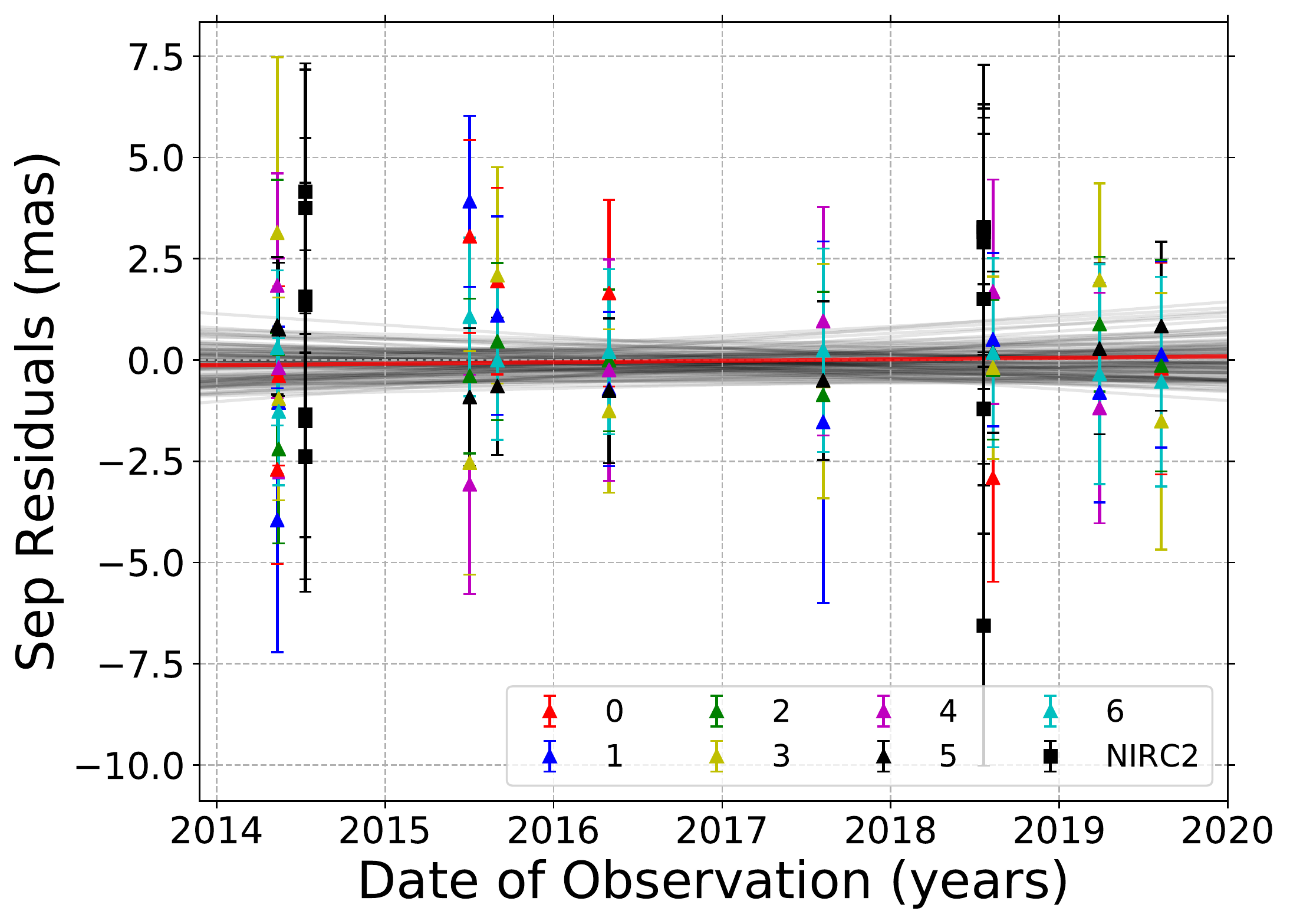}
\endminipage\hfill
\minipage{0.5\textwidth}
  \includegraphics[width=\linewidth]{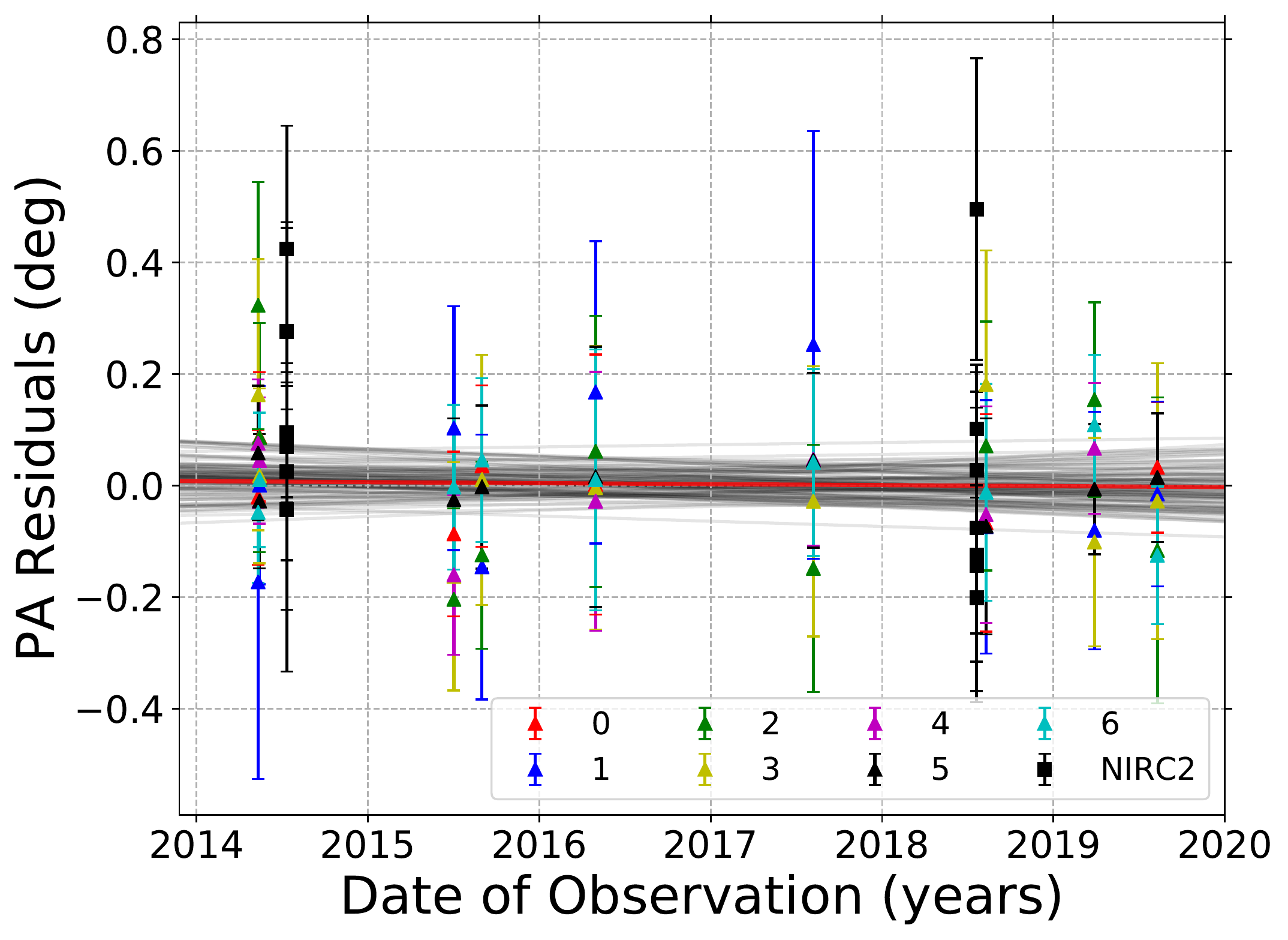}
\endminipage
    \caption{Residuals of each of the background stars in RA (\textbf{Top Left}), Dec (\textbf{Top Right}), separation (\textbf{Bottom Left}), and position angle (\textbf{Bottom Right}), with a corresponding distribution of linear fits to the residuals overplotted as gray lines. The ``best-fit" linear model was generated using the median values of the MCMC posterior distributions and is plotted as the red line. Numbers 0-6 in the legend label the GPI residuals for the 7 background stars (plotted as colored triangles). The two epochs of NIRC2 residuals are plotted as black squares to differentiate from the GPI measurements. The NIRC2 residuals were explicitly excluded from the linear fit. See Fig. \ref{fig:weighted_residuals} for similar plots but using the weighted mean of the residuals instead for better readability. \label{fig:residuals}}
\end{figure*}

\begin{figure*}
    \centering
\minipage{0.5\textwidth}
  \includegraphics[width=\linewidth]{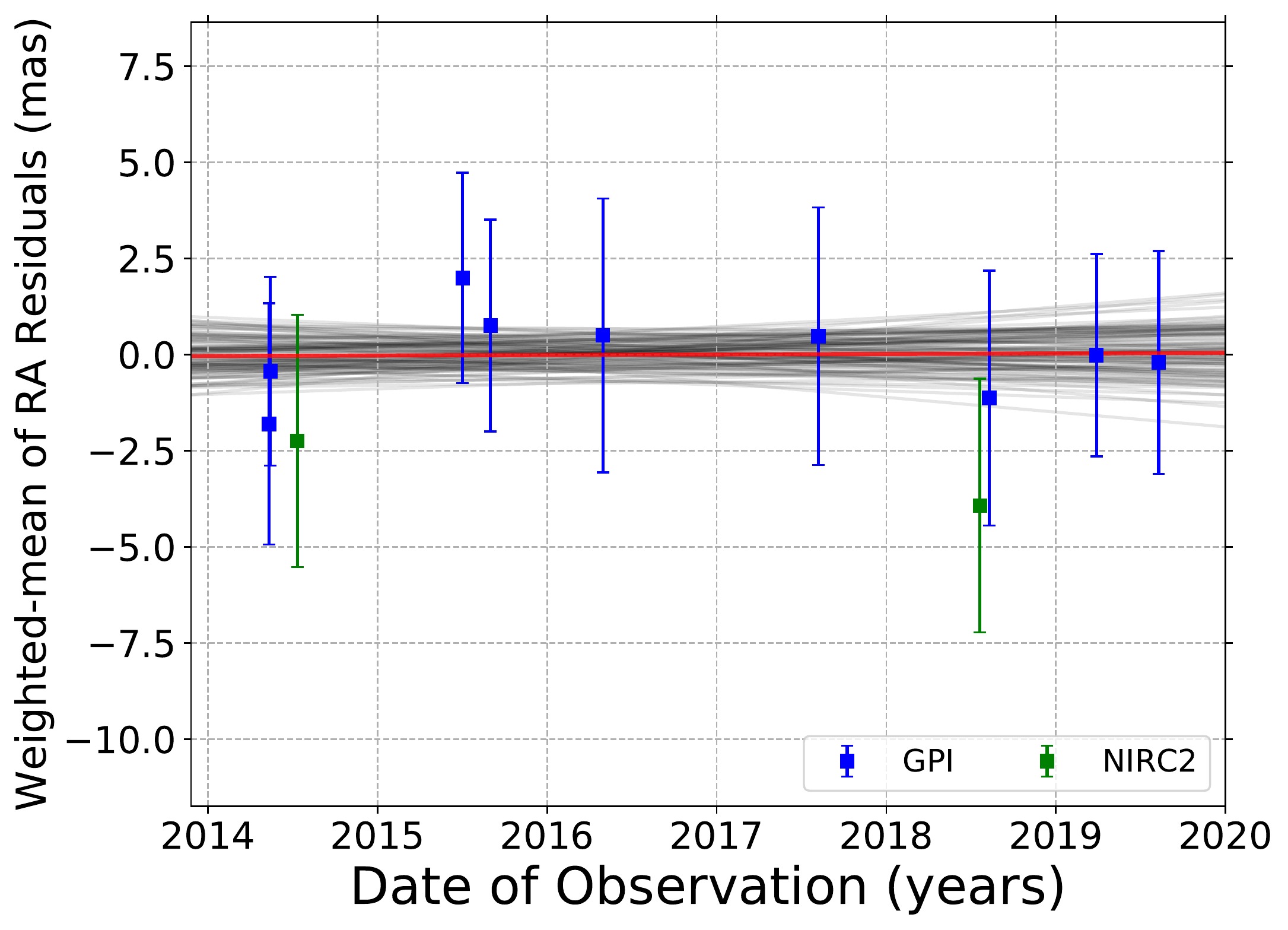}
\endminipage\hfill
\minipage{0.5\textwidth}
  \includegraphics[width=\linewidth]{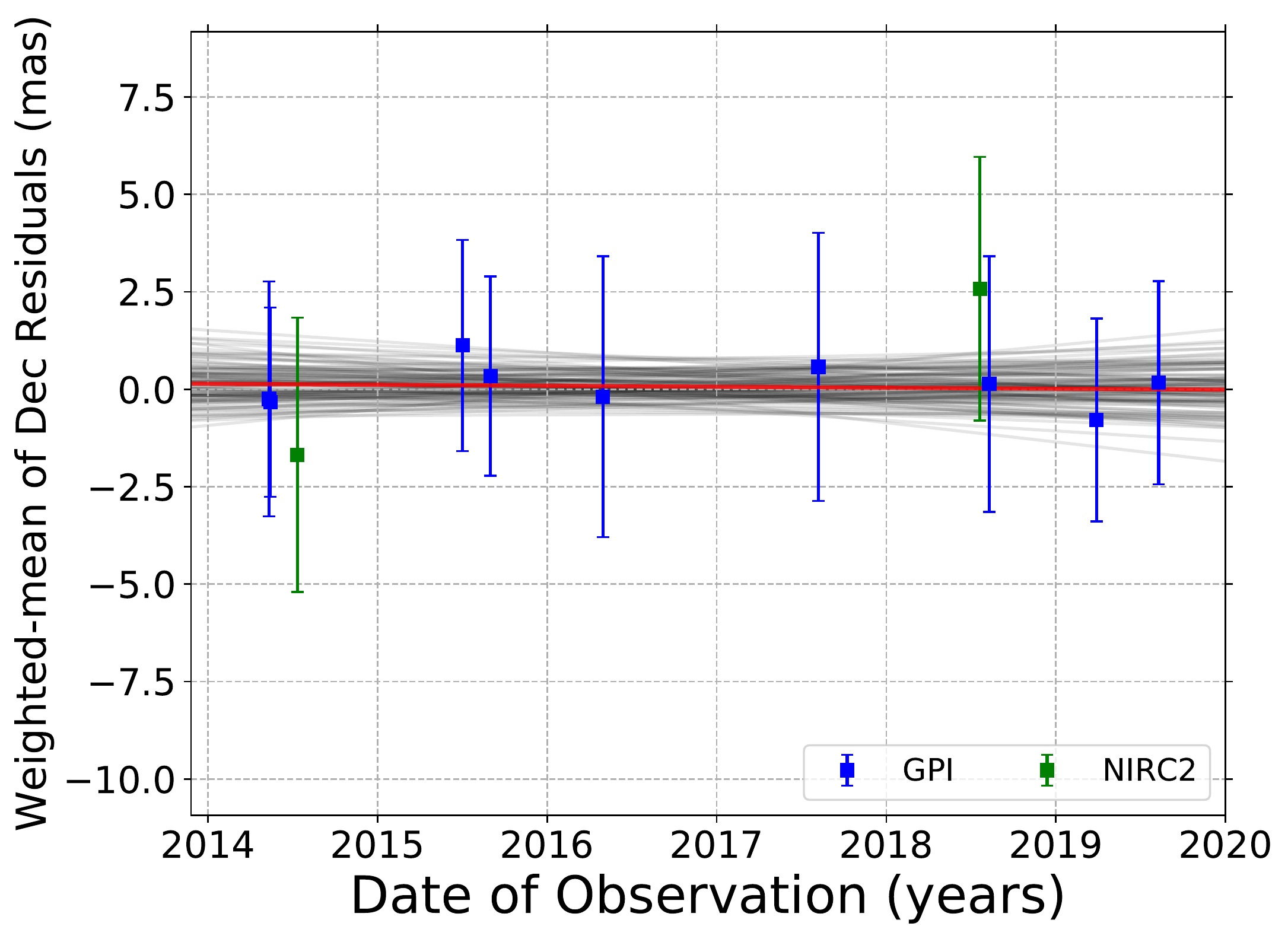}
\endminipage\hfill
\minipage{0.5\textwidth}
  \includegraphics[width=\linewidth]{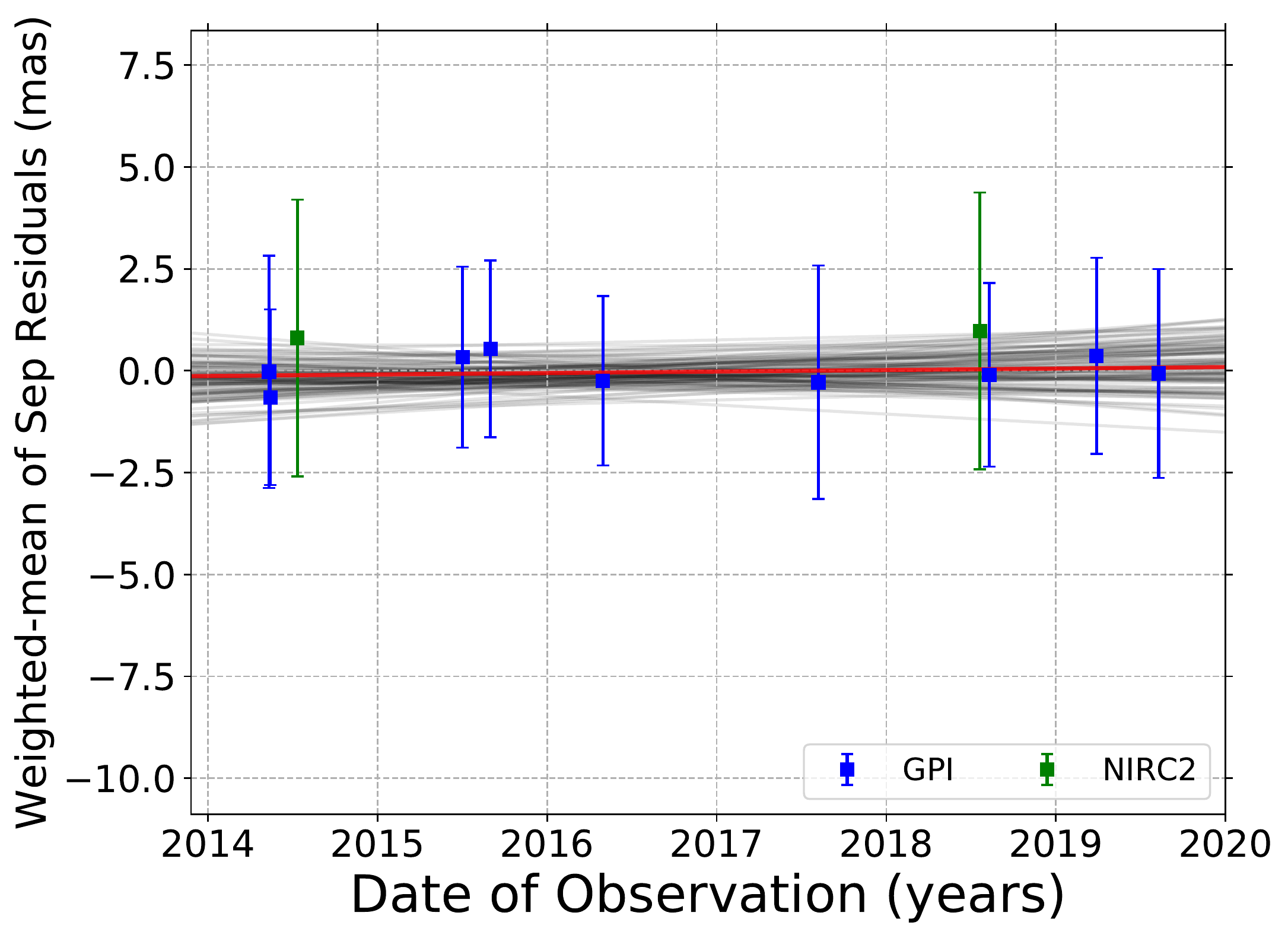}
\endminipage\hfill
\minipage{0.5\textwidth}
  \includegraphics[width=\linewidth]{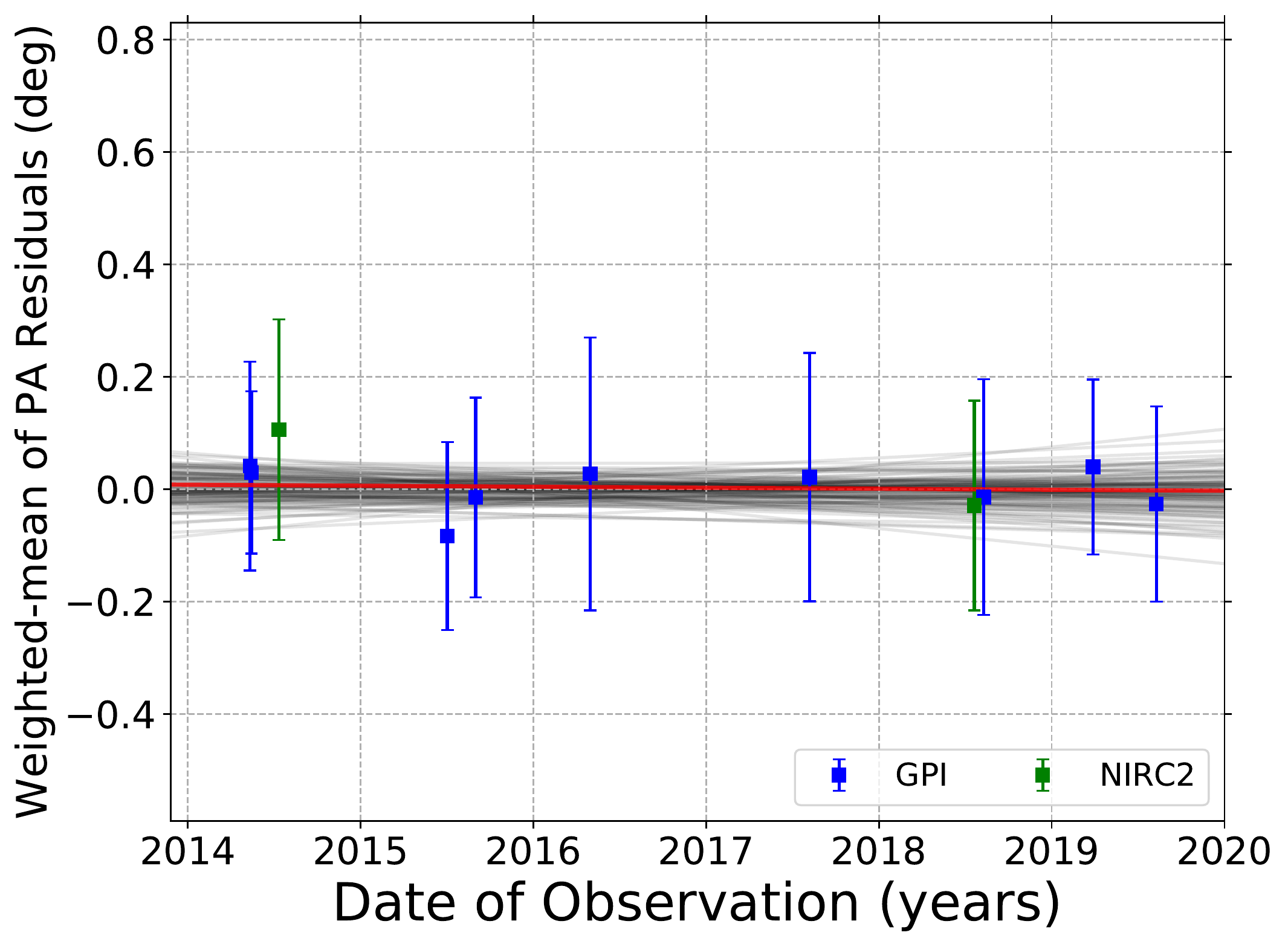}
\endminipage
    \caption{Weighted-mean of the residuals in RA (\textbf{Top Left}), Dec (\textbf{Top Right}), separation (\textbf{Bottom Left}), and position angle (\textbf{Bottom Right}), with a corresponding distribution of linear fits to the residuals overplotted as gray lines. This is essentially the same plot as Fig. \ref{fig:residuals} but using the weighted-mean of the residuals instead for better readability. The weighted-mean was calculated using the inverse square of the error bar for each data point as weights. The error bars associated with each weighted-mean were simply taken to be the average of the error bars of the data points in a given epoch.\label{fig:weighted_residuals}}
\end{figure*}

We produced a sample of 100 model curves by sampling 100 times from the posterior distributions for the MCMC parameters. We then took the mean of these curves to produce a best-fit model curve. Finally, we subtracted the measured position of the given background star at a given epoch from the best-fit model curve's predicted position at that given epoch (the error bars on the residuals are carried over from BKA), producing the residuals that we used to look for calibration systematics and offsets in GPI (see Fig.~\ref{fig:residuals}). We produced similar best-fit model curves and residuals in separation and position angle by doing the appropriate coordinate transformations on the measured positions in RA/Dec. We also calculated the weighted mean of the residuals at each epoch (Fig.~\ref{fig:weighted_residuals}), using $1/\sigma^2$ as the weight of each data point. We adopted the average uncertainty of the residuals at each epoch as the error bar; this better captures the systematic errors than the standard error that assumes independent measurements but leads to an overestimate of the magnitude of the uncertainty.
    
We can put a numerical constraint on possible drifts in the astrometric calibration of GPI over the approximately five year temporal baseline of observations by measuring the rate of change of the residuals as a function of time. We fit the residuals with a simple linear fit model using an MCMC-based approach, with the gradient and y-intercept of the fit (parameters $m$ and $b$ respectively in the traditional notation of a line $y = mx + b$) being drawn from uniform priors. 

\startlongtable
\begin{deluxetable*}{ccccc}
\tablecaption{Best-fit linear model parameter values for GPI residuals\label{tbl:linear_fit}}
\tablehead{\colhead{} & \colhead{RA} & \colhead{Dec} & \colhead{Separation} & \colhead{Position Angle}}
\startdata
Gradient & $-0.02_{-0.18}^{+0.18}$ mas/yr& $-0.03_{-0.18}^{+0.17}$ mas/yr& $0.04_{-0.16}^{+0.16}$ mas/yr & $0.002_{-0.010}^{+0.010}$ deg/yr\\
Intercept \tablenotemark{*} & $-0.04_{-0.52}^{+0.54}$ mas & $0.15_{-0.52}^{+0.51}$ mas & $-0.13_{-0.45}^{+0.46}$ mas & $0.008_{-0.032}^{+0.031}$ deg \\
\enddata
\tablenotetext{*}{The intercept is set at the first epoch of observations: 2014 May 12}
\end{deluxetable*}

As can be seen in Fig. \ref{fig:residuals}, the residuals for the GPI measurements in all coordinates (RA, Dec, separation, position angle) imply that the astrometric calibration of GPI has been stable over the five year baseline of observations. The best-fit linear model of the residuals has a slope consistent with zero for all coordinates. It is particularly informative to look specifically at the residuals in separation and position angle (PA) because these coordinates would reveal any systematics in plate scale and true north angle. GPI's plate scale and true north angle both seem to be stable over time. Fig. \ref{fig:residuals} was useful for identifying the issue with the cassegrain rotator described in Sec. \ref{sec:parallactic angle correction}. The PA residuals shown in Fig. \ref{fig:residuals} are with the correction for the Cassegrain rotator offset applied.

The NIRC2 data was not included in our best-fit model curves generated using MCMC (see Sec. \ref{sec:parameter estimation}). This separate dataset was obtained to serve as a secondary check of the validity of the astrometric solution obtained by GPI. By comparing the residuals of the predicted GPI model curves with the measured background star positions in the NIRC2 data, we could look for  discrepancies that might point to astrometric calibration errors in GPI. The residuals in separation angle and position angle appear to be consistent between instruments, indicating that GPI's plate scale and true north angle is well calibrated relative to NIRC2.

\subsection{Background star relative positions}
The orientation and magnification of GPI's astrometric frame relative to NIRC2 was also measured by comparing the relative separation and position angles between all unique pairs of background stars within the near-contemporaneous epochs obtained in 2014 and 2018. We did this comparison as a secondary check of the validity of the residuals generated in Sec. \ref{sec:residuals}. The detector positions of each of the seven background stars within the final PSF-subtracted image measured using BKA (Sec.~\ref{sec:klip}) were used to calculate separations and position angles between the 21 unique pairs of stars in detector coordinates. These relative measurements incorporated the nominal true north offsets for both GPI (given in Table~\ref{tbl:log}) and NIRC2 ($0.252$ or $0.262$ depending on epoch, \citealp{Yelda:2010ig,2016PASP..128i5004S}). As the relative offsets between pairs of background stars did not require the foreground star center location, which was a significant source of uncertainty in the final astrometry of the background stars presented in Table~\ref{tbl:astrometry}, the relative separations and position angles could be measured more precisely.

\begin{figure}
\includegraphics[width=\columnwidth]{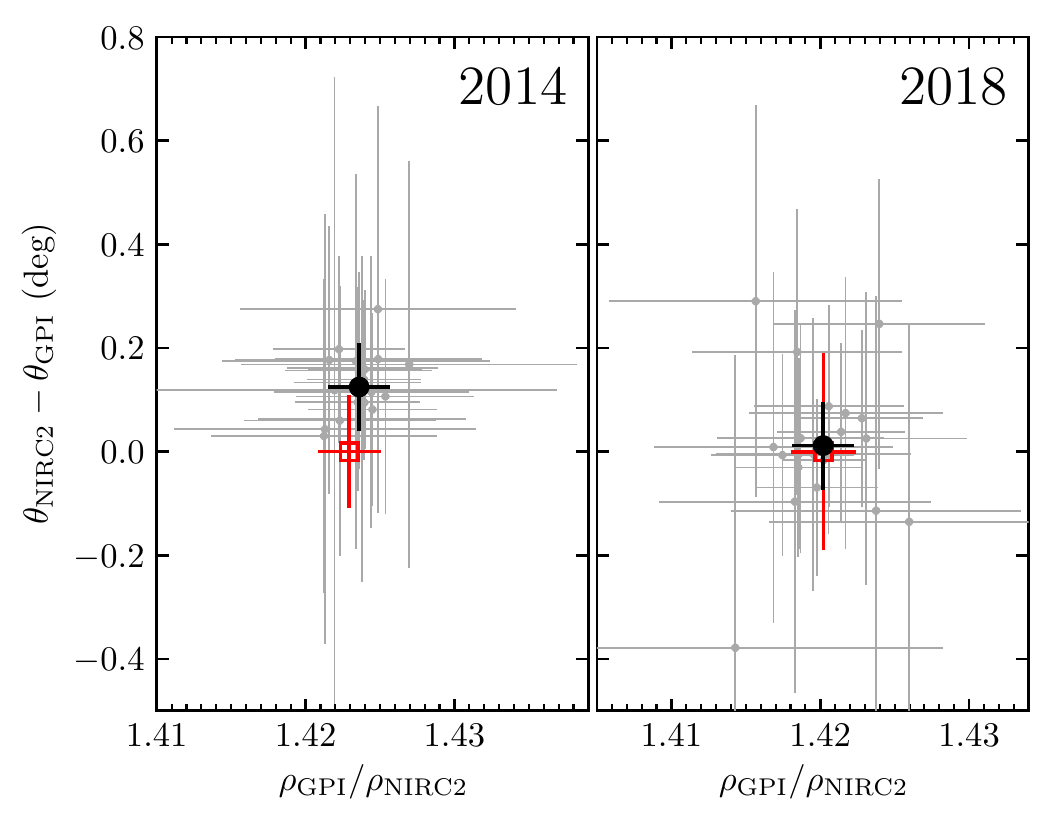}
\caption{Ratio of plate scales and difference in position angles for GPI and NIRC2 computed from the positions of the seven background stars relative to one another in 2014 (left) and 2018 (right). The fitted offset and ratio (black circle) and corresponding nominal values (red open square) are also shown. A GPI observation of the position of the background stars was simulated at the epoch of the NIRC2 observations using the model described in Section~\ref{sec:astrometry}.}
\label{fig:relative_fit}
\end{figure}

Using the relative positions of the unique pairs of background objects for both instruments at both epochs, we performed a $\chi^2$ minimization to find the magnification factor ($\rho_{\rm GPI}/\rho_{\rm NIRC2}$) and rotation ($\theta_{\rm NIRC2} - \theta_{\rm GPI}$) that would best align the measured offsets from both instruments at a given epoch. We found a magnification factor of $1.4245\pm0.0015$ (cf $1.4229\pm0.0021$ using the nominal plate scales) and a rotation angle of $0\fdg077\pm0\fdg062$\ for the 2014 epoch, and a magnification factor of $1.4200\pm0.0015$ (cf $1.4202\pm0.0022$, similarly) and rotation angle of $0\fdg021\pm0\fdg062$ for the 2018 epoch. We found a minimum $\chi^2_{\nu}$ of 0.81 for the 2014 epoch and 0.83 for the 2018 epoch. The magnification factor was consistent for both epochs, however the rotation angle was marginally consistent for the 2014 epoch (1.1 $\sigma$), but consistent for the 2018 epoch. The small offset in the rotation angle seen for the 2014 epoch is consistent with the position angle residuals seen in Figure~\ref{fig:residuals}.

The non-zero motion of the background objects relative to one another was a potential source of bias when comparing measurements from the two instruments that were not obtained on the same date. The offset between the GPI and NIRC2 measurements was 58 days in 2014 and 19 days in 2018. To account for the non-zero relative motion of the background stars, we repeated the same fit but instead of using the GPI measurements nearest in time to the NIRC2 observations, we used the model described in Sec.~\ref{sec:astrometry} to predict the locations of the objects within a hypothetical GPI observation taken on the same date. We found a magnification factor of $1.4239\pm0.0021$ and a rotation angle of $0\fdg1316\pm0\fdg086$ for the 2014 epoch, and a magnification factor of $1.4197\pm0.0021$ and rotation angle of $0\fdg004\pm0\fdg085$ for the 2018 epoch. We found a minimum $\chi^2_{\nu}$ of 0.12 for the 2014 epoch and 0.31 for the 2018 epoch, lower than previously due to the larger uncertainties on the positions of the background stars in the simulated GPI measurement at the NIRC2 epoch. As with the previous comparison, the two plate scales are consistent with the nominal values, but a small (1.5-$\sigma$) offset is measured for the rotation between the 2014 NIRC2 epoch and the simulated GPI measurement. This offset is not significant when the uncertainty in the north offset angle for both instruments is included, decreasing to a 0.9-$\sigma$ discrepancy. This analysis, and that previously discussed in Section~\ref{sec:residuals}, demonstrates the consistency between the astrometric solution of NIRC2 and GPI.

\section{Conclusion/Future Work} \label{sec:conclusion}
Using high-contrast direct imaging and PSF subtraction, a five parameter astrometric model, and MCMC fitting techniques, we have characterized the positions, and parallactic and proper motions of seven sources in the field of view around HD 165054. The measured astrometry confirms our initial assumption that these sources are distant galactic bulge stars, making them good candidates for follow-up observations for the purpose of re-calibration or comparison of astrometric calibration with other high-contrast direct imagers. An analysis of the residuals of our five parameter astrometric model confirms the stability of GPI's calibration as well as the consistency of the plate scale and true north offset of the instrument relative to the well-calibrated NIRC2 instrument on Keck. 

There is potential for follow-up GPI or NIRC2 observations of HD 165054 to further constrain the astrometry of the sources in the field. In addition to the two epochs of NIRC2 data, there is potentially available SPHERE IRDIS data of HD 165054 that would be interesting to include in our analysis to see whether the astrometry obtained between GPI and SPHERE is consistent. Analysis of IRDIS data of HD 165054 could help diagnose inconsistencies in measured astrometry (most notably differences in measured PA) that have been observed between GPI and SPHERE in other systems \citep{Maire:2019}.

This study was done using seven of the brightest background sources in the field of view. There are certainly more than seven background sources we could have used (See Fig. \ref{fig:pyklip}), but we only chose to do seven for now because the high SNR of the brightest sources gives us the smallest error bars on our astrometry. More sophisticated source identification algorithms could be used to include additional background stars in our fit (e.g., \citealp{2017ApJ...842...14R}), but the lower SNR of these stars probably would not lead to a significant improvement in our characterization of GPI's astrometric solution. 
 
Baade's Window has been a known region of low extinction from interstellar dust for many decades, and it has served astronomers well in studying the properties of the central bulge stars of our galaxy. With the need for precision astrometry in order to probe key properties of directly imaged exoplanets and with the specific calibration requirements of the Gemini Planet Imager, HD 165054 is a powerful tool in understanding the instrument's astrometric solution. We hope that further observation of HD 165054 by GPI and other high-contrast imagers will improve the precision of exoplanet astrometry and our understanding of calibration systematics of the instruments we use.

\acknowledgements

This work is based on observations obtained at the Gemini Observatory, which is operated by the Association of Universities for Research in Astronomy, Inc. (AURA), under a cooperative agreement with the National Science Foundation (NSF) on behalf of the Gemini partnership: the NSF (United States), the National Research Council (Canada), CONICYT (Chile), Ministerio de Ciencia, Tecnolog\'ia e Innovaci\'on Productiva (Argentina), and Minist\'erio da Ci\^encia, Tecnologia e Inova\c c\~ao (Brazil). Some of the data presented herein were obtained at the W.~M.~Keck Observatory, which was made possible by the generous financial support of the W.~M.~Keck Foundation and is operated as a scientific partnership among the California Institute of Technology, the University of California, and NASA. The authors wish to recognize and acknowledge the very significant cultural role and reverence that the summit of Mauna Kea has always had within the indigenous Hawaiian community. We are most fortunate to have the opportunity to conduct observations from this mountain. This work has also made use of data from the European Space Agency mission {\it Gaia} (\url{https://www.cosmos.esa.int/gaia}), processed by the {\it Gaia} Data Processing and Analysis Consortium (DPAC, \url{https://www.cosmos.esa.int/web/gaia/dpac/consortium}). Funding for the DPAC has been provided by national institutions, in particular the institutions participating in the {\it Gaia} Multilateral Agreement. This research has made use of the SIMBAD database and the VizieR catalog access tool, both operated at the CDS, Strasbourg, France. This research has made use of the Washington Double Star Catalog maintained at the U.S. Naval Observatory.

Supported by NSF grants AST-1411868 (E.L.N., K.B.F., B.M., and J.P.), AST-141378 (G.D.), and AST-1518332 (M.M.N., R.J.D.R., T.M.E., J.R.G., P.K., G.D.). Supported by NASA grants NNX14AJ80G (E.L.N., B.M., F.M., and M.P.), NNX15AC89G and NNX15AD95G/NExSS (M.M.N., R.J.D.R., B.M., T.M.E., J.J.W., G.D., J.R.G., P.K.), and NN15AB52l (D.S.). M.M.B. was supported by NASA through Hubble Fellowship grant \#51378.01-A, awarded by the Space Telescope Science Institute, which is operated by AURA, for NASA, under contract NAS5-26555. K.W.D. is supported by an NRAO Student Observing Support Award SOSPA3-007. J.J.W. is supported by the Heising-Simons Foundation 51 Pegasi b postdoctoral fellowship. This work benefited from NASA's Nexus for Exoplanet System Science (NExSS) research coordination network sponsored by NASA's Science Mission Directorate.

\software{Gemini Planet Imager Data Pipeline (\citealt{Perrin:2014jh, Perrin:2016}, \url{http://ascl.net/1411.018}), pyKLIP (\citealt{Wang:2015pyklip}, \url{http://ascl.net/1506.001}), emcee (\citealt{ForemanMackey:2013io}, \url{http://ascl.net/1303.002}), Astropy \citep{TheAstropyCollaboration:2013cd}.}

\facilities{Gemini:South, Keck:II (NIRC2)}

\bibliographystyle{apj}   
\bibliography{refs}

\appendix

\section{Measured positions}
\label{sec:measured positions}

We include a table of the measured positions in RA and Dec of each the seven background stars characterized in this study relative to HD 165054. Each background star should have eleven total epochs of observations (9 using GPI and 2 using NIRC2). The NIRC2 epochs have been differentiated from the GPI epochs using a superscript marker.

\startlongtable
\begin{deluxetable}{cccc}
\tablecaption{Measured offset between HD 165054 and the seven background stars for each epoch \label{tbl:astrometry}}
\tablehead{\colhead{Star} & \colhead{Date} & \colhead{$\Delta \alpha^{\star}$ (mas)} &  \colhead{$\Delta \delta$ (mas)}}
 \startdata
\hline
0 & 2014-05-12 & $1078.53\pm2.58$ & $843.63\pm2.67$\\
& 2014-05-15 & $1082.44\pm2.44$ & $843.76\pm2.57$\\
& 2014-07-12\tablenotemark{a} & $1094.75\pm3.02$ & $856.49\pm3.03$\\
& 2015-07-03 & $1060.8\pm3.06$ & $942.77\pm3.12$\\
& 2015-09-01 & $1071.49\pm2.96$ & $952.31\pm3.1$\\
& 2016-04-30 & $1015.96\pm4.43$ & $1007.26\pm4.43$\\
& 2017-08-07 & \nodata & \nodata\\
& 2018-07-21\tablenotemark{a} & $962.03\pm3.07$ & $1191.69\pm3.07$\\
& 2018-08-10 & $967.22\pm4.39$ & $1192.28\pm3.83$\\
& 2019-03-29 & \nodata & \nodata\\
& 2019-08-10 & $938.64\pm3.07$ & $1274.72\pm2.81$\\
\hline
1 & 2014-05-12 & $554.96\pm3.68$ & $346.8\pm3.67$\\
& 2014-05-15 & $558.97\pm1.96$ & $347.41\pm1.96$\\
& 2014-07-12\tablenotemark{a} & $572.64\pm4.07$ & $359.38\pm4.86$\\
& 2015-07-03 & $546.07\pm2.54$ & $439.1\pm2.25$\\
& 2015-09-01 & $552.07\pm2.79$ & $451.68\pm2.64$\\
& 2016-04-30 & $500.7\pm2.84$ & $499.82\pm2.59$\\
& 2017-08-07 & $493.47\pm4.64$ & $598.9\pm5.03$\\
& 2018-07-21\tablenotemark{a} & $458.98\pm3.88$ & $679.66\pm3.97$\\
& 2018-08-10 & $462.63\pm3.03$ & $682.44\pm2.46$\\
& 2019-03-29 & $410.01\pm3.04$ & $729.68\pm2.78$\\
& 2019-08-10 & $434.07\pm2.54$ & $760.23\pm2.3$\\
\hline
2 & 2014-05-12 & $616.19\pm4.14$ & $-767.13\pm3.28$\\
& 2014-05-15 & $617.89\pm3.2$ & $-761.54\pm2.77$\\
& 2014-07-12\tablenotemark{a} & $629.84\pm3.55$ & $-754.77\pm4.13$\\
& 2015-07-03 & $607.5\pm2.36$ & $-666.65\pm2.17$\\
& 2015-09-01 & $617.66\pm2.37$ & $-655.62\pm2.25$\\
& 2016-04-30 & $564.04\pm2.87$ & $-602.9\pm2.66$\\
& 2017-08-07 & $560.29\pm2.75$ & $-495.87\pm2.71$\\
& 2018-07-21\tablenotemark{a} & $524.99\pm3.36$ & $-414.96\pm3.51$\\
& 2018-08-10 & $532.69\pm2.13$ & $-416.06\pm2.3$\\
& 2019-03-29 & $482.53\pm1.75$ & $-367.01\pm1.77$\\
& 2019-08-10 & $507.1\pm2.94$ & $-332.71\pm2.57$\\
\hline
3 & 2014-05-12 & $-413.17\pm4.52$ & $-943.92\pm4.19$\\
& 2014-05-15 & $-408.5\pm2.76$ & $-940.42\pm2.55$\\
& 2014-07-12\tablenotemark{a} & $-400.2\pm3.27$ & $-929.16\pm3.58$\\
& 2015-07-03 & $-416.94\pm3.28$ & $-844.41\pm2.87$\\
& 2015-09-01 & $-410.65\pm3.47$ & $-834.43\pm2.89$\\
& 2016-04-30 & $-458.05\pm3.63$ & $-775.67\pm2.61$\\
& 2017-08-07 & $-459.64\pm3.37$ & $-669.05\pm2.96$\\
& 2018-07-21\tablenotemark{a} & $-494.92\pm3.64$ & $-586.99\pm3.98$\\
& 2018-08-10 & $-485.54\pm2.91$ & $-582.73\pm2.6$\\
& 2019-03-29 & $-532.8\pm2.52$ & $-534.85\pm2.32$\\
& 2019-08-10 & $-506.51\pm3.29$ & $-499.43\pm2.91$\\
\hline
4 & 2014-05-12 & $-1569.34\pm2.94$ & $-763.93\pm3.36$\\
& 2014-05-15 & $-1566.59\pm2.87$ & $-763.14\pm3.31$\\
& 2014-07-12\tablenotemark{a} & $-1557.74\pm3.01$ & $-750.59\pm3.01$\\
& 2015-07-03 & $-1580.8\pm3.01$ & $-673.83\pm4.08$\\
& 2015-09-01 & \nodata & \nodata\\
& 2016-04-30 & $-1628.56\pm3.55$ & $-604.05\pm6.65$\\
& 2017-08-07 & $-1640.34\pm3.04$ & $-497.17\pm4.5$\\
& 2018-07-21\tablenotemark{a} & $-1676.18\pm3.08$ & $-417.24\pm3.06$\\
& 2018-08-10 & $-1670.3\pm3.05$ & $-417.24\pm5.69$\\
& 2019-03-29 & $-1720.9\pm2.9$ & $-362.34\pm3.55$\\
& 2019-08-10 & \nodata & \nodata\\
\hline
5 & 2014-05-12 & $-898.46\pm1.76$ & $364.02\pm1.99$\\
& 2014-05-15 & $-898.4\pm1.71$ & $363.4\pm1.99$\\
& 2014-07-12\tablenotemark{a} & $-885.57\pm3.01$ & $377.38\pm3.02$\\
& 2015-07-03 & $-914.76\pm1.92$ & $459.55\pm2.46$\\
& 2015-09-01 & $-904.75\pm1.93$ & $473.05\pm2.44$\\
& 2016-04-30 & $-957.22\pm2.66$ & $529.7\pm3.99$\\
& 2017-08-07 & $-965.27\pm2.4$ & $638.71\pm2.85$\\
& 2018-07-21\tablenotemark{a} & $-1000.89\pm3.04$ & $721.81\pm3.03$\\
& 2018-08-10 & $-996.07\pm2.93$ & $722.66\pm3.56$\\
& 2019-03-29 & $-1047.27\pm2.32$ & $776.25\pm2.47$\\
& 2019-08-10 & $-1024.69\pm2.32$ & $809.53\pm2.43$\\
\hline
6 & 2014-05-12 & $-236.41\pm2.35$ & $1046.76\pm1.93$\\
& 2014-05-15 & $-234.47\pm2.24$ & $1046.2\pm1.84$\\
& 2014-07-12\tablenotemark{a} & $-220.74\pm3.01$ & $1058.57\pm3.03$\\
& 2015-07-03 & $-253.17\pm2.97$ & $1141.4\pm2.01$\\
& 2015-09-01 & $-242.11\pm2.98$ & $1153.03\pm2.01$\\
& 2016-04-30 & $-294.98\pm4.96$ & $1207.27\pm2.31$\\
& 2017-08-07 & $-302.73\pm3.89$ & $1311.59\pm2.59$\\
& 2018-07-21\tablenotemark{a} & $-340.12\pm3.05$ & $1391.44\pm3.06$\\
& 2018-08-10 & $-332.62\pm4.76$ & $1393.38\pm2.53$\\
& 2019-03-29 & $-380.85\pm3.27$ & $1443.93\pm2.72$\\
& 2019-08-10 & $-364.5\pm3.24$ & $1473.51\pm2.62$\\
\enddata
\tablenotetext{a}{Keck/NIRC2 measurements}
\end{deluxetable}

\section{Best-fit Model and Residuals for Bayesian KLIP-FM Astrometry}

We include the data stamps, best-fit models, and corresponding residuals obtained when characterizing the astrometry of the stars  in this study. 

\begin{figure*}
    \centering
\minipage{0.50\textwidth}
  \includegraphics[width=\linewidth]{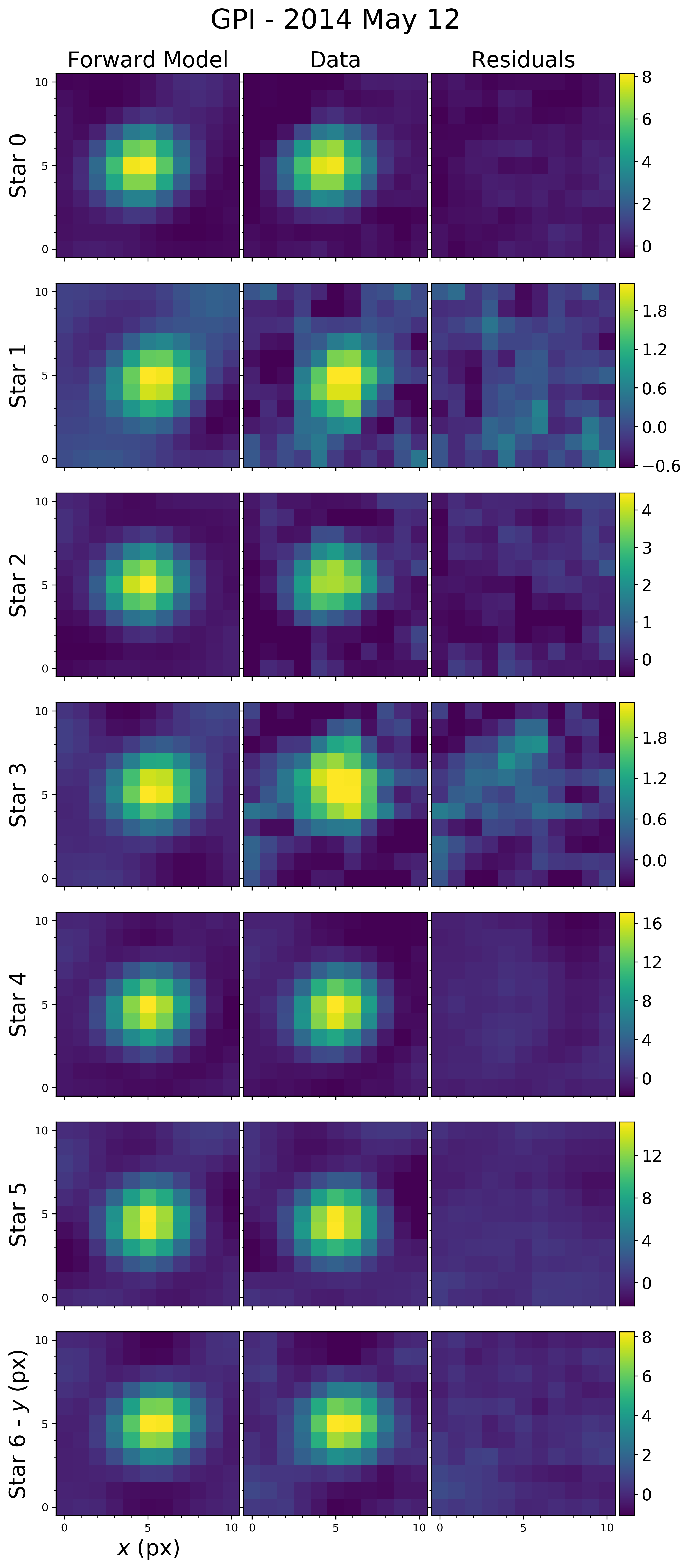}
\endminipage\hfill
\minipage{0.50\textwidth}
  \includegraphics[width=\linewidth]{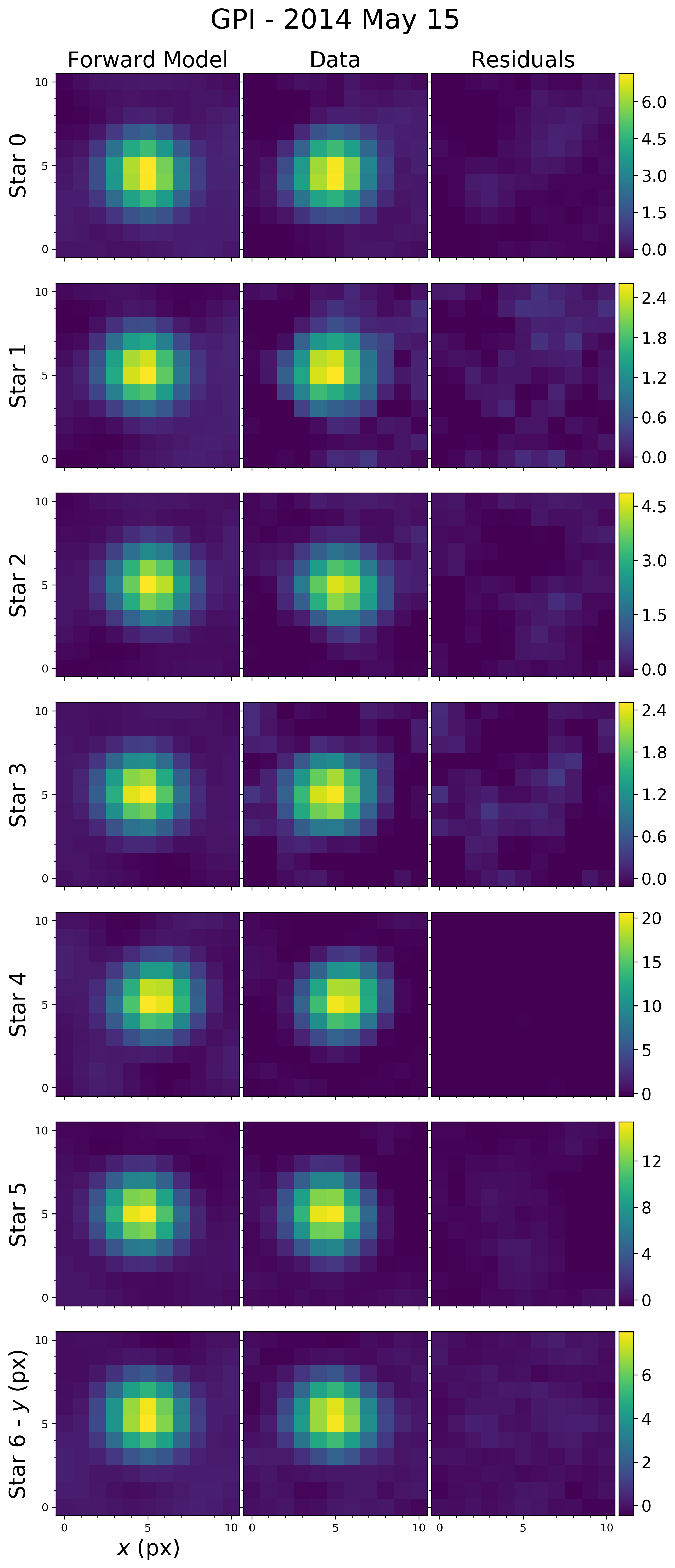}
\endminipage
\end{figure*}

\begin{figure*}
    \centering
\minipage{0.50\textwidth}
  \includegraphics[width=\linewidth]{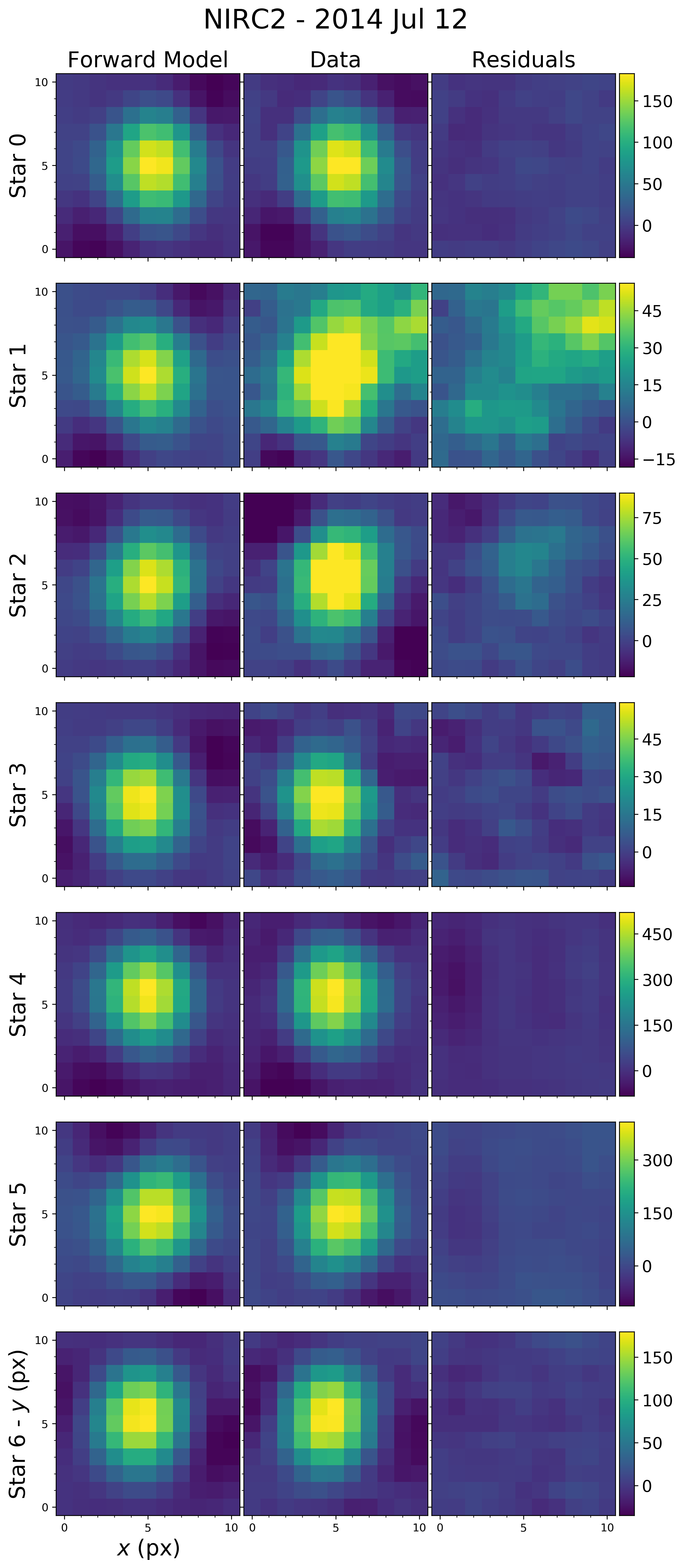}
\endminipage\hfill
\minipage{0.50\textwidth}
  \includegraphics[width=\linewidth]{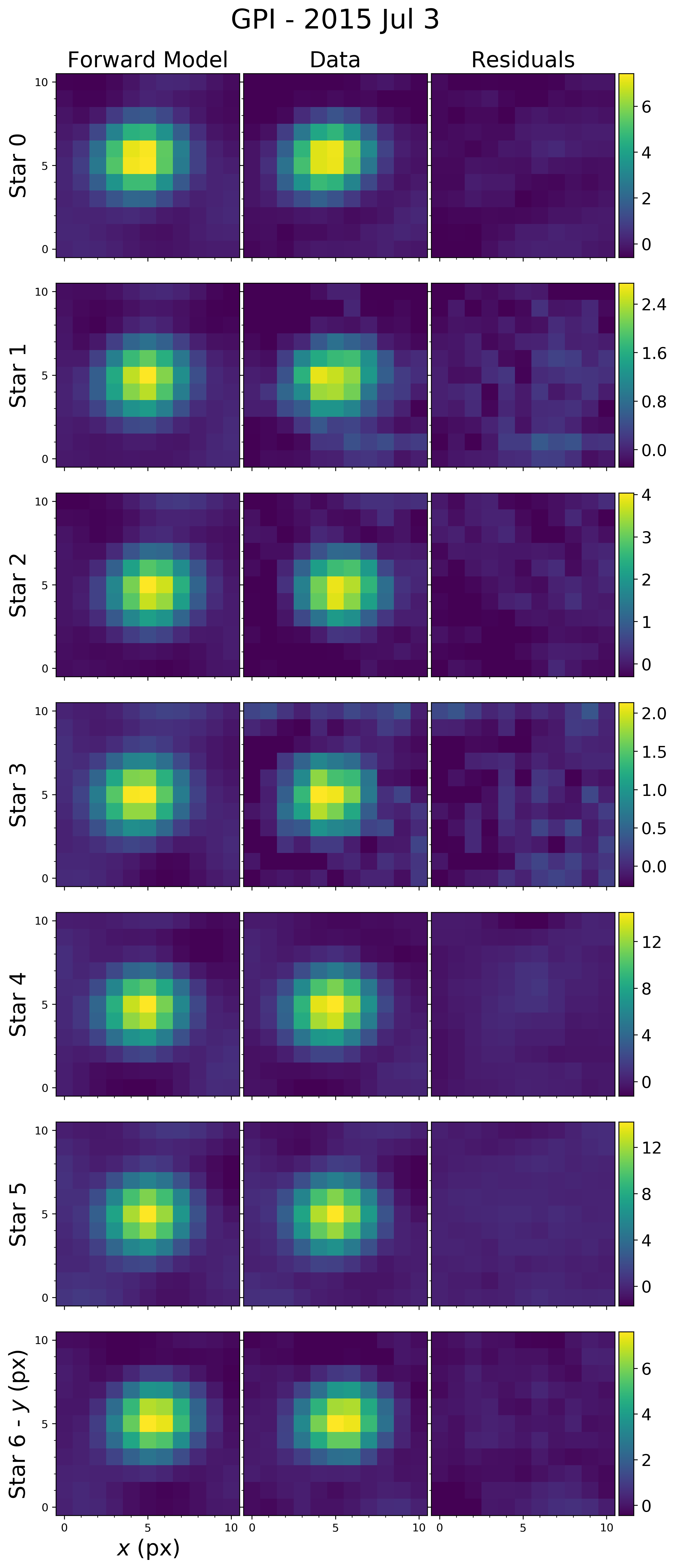}
\endminipage
\end{figure*}

\begin{figure*}
    \centering
\minipage{0.50\textwidth}
  \includegraphics[width=\linewidth]{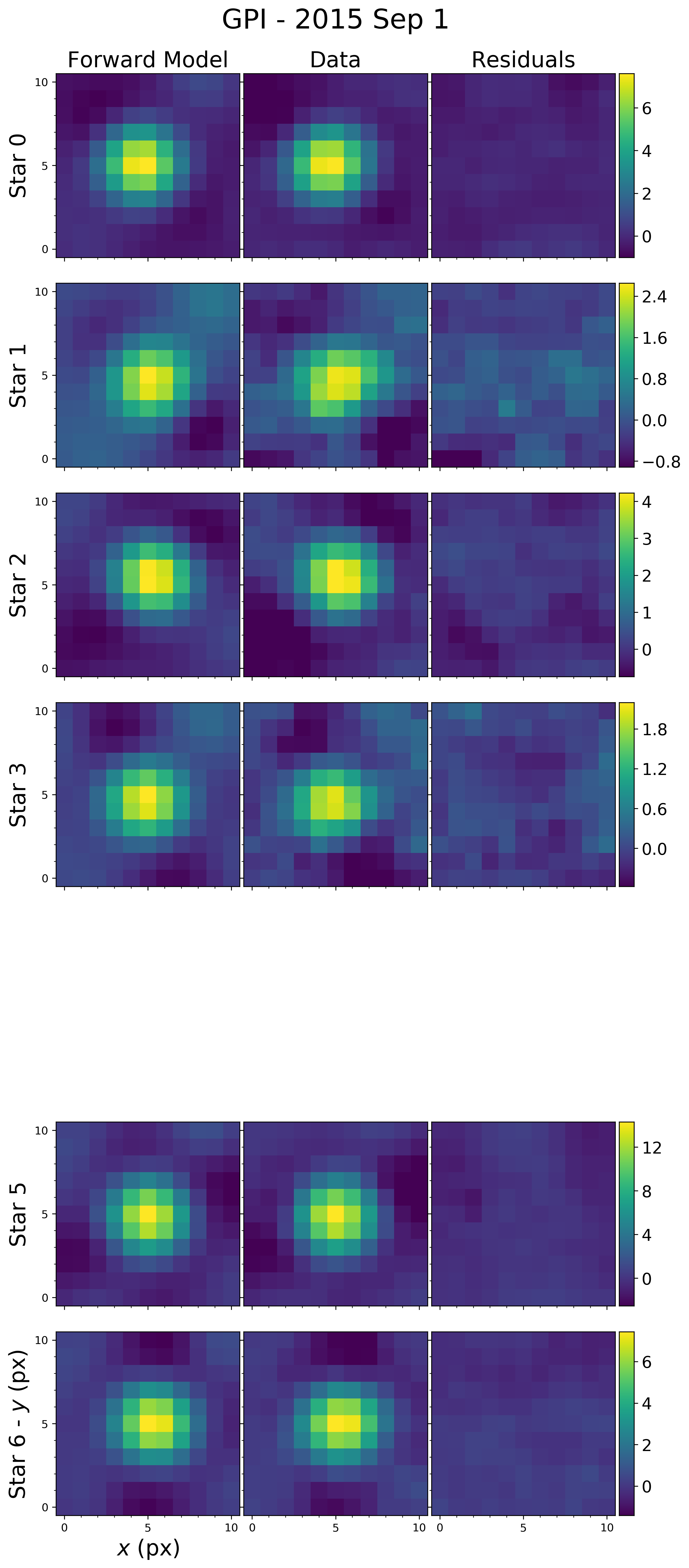}
\endminipage\hfill
\minipage{0.50\textwidth}
  \includegraphics[width=\linewidth]{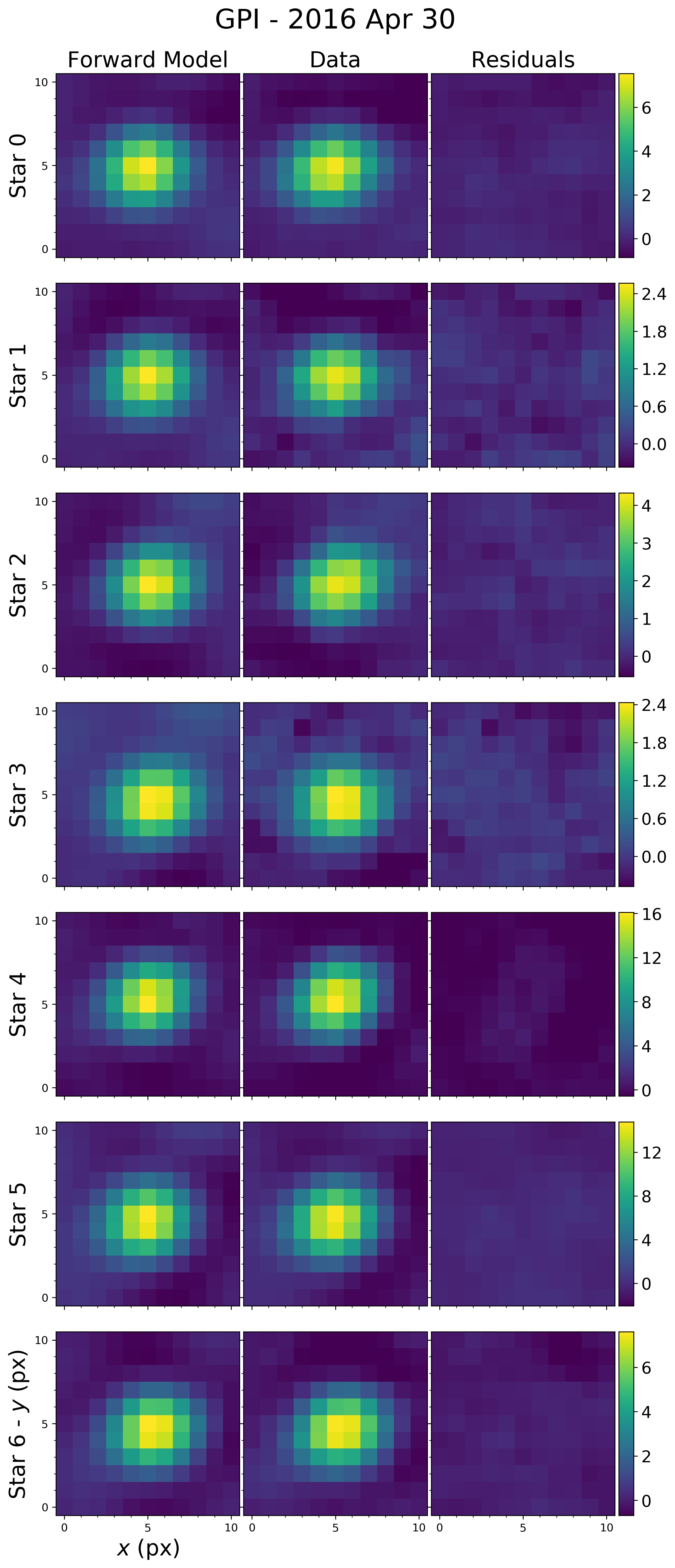}
\endminipage
\end{figure*}

\begin{figure*}
    \centering
\minipage{0.50\textwidth}
  \includegraphics[width=\linewidth]{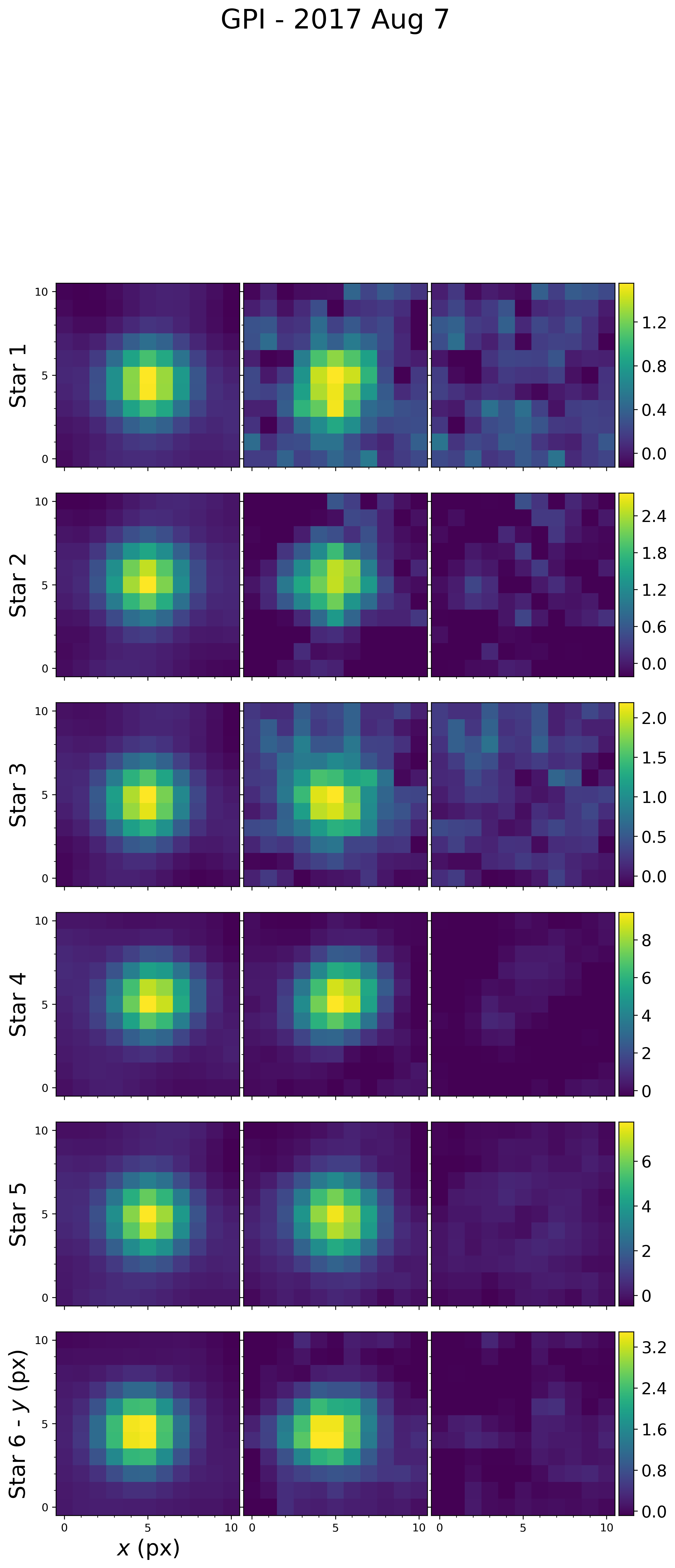}
\endminipage\hfill
\minipage{0.50\textwidth}
  \includegraphics[width=\linewidth]{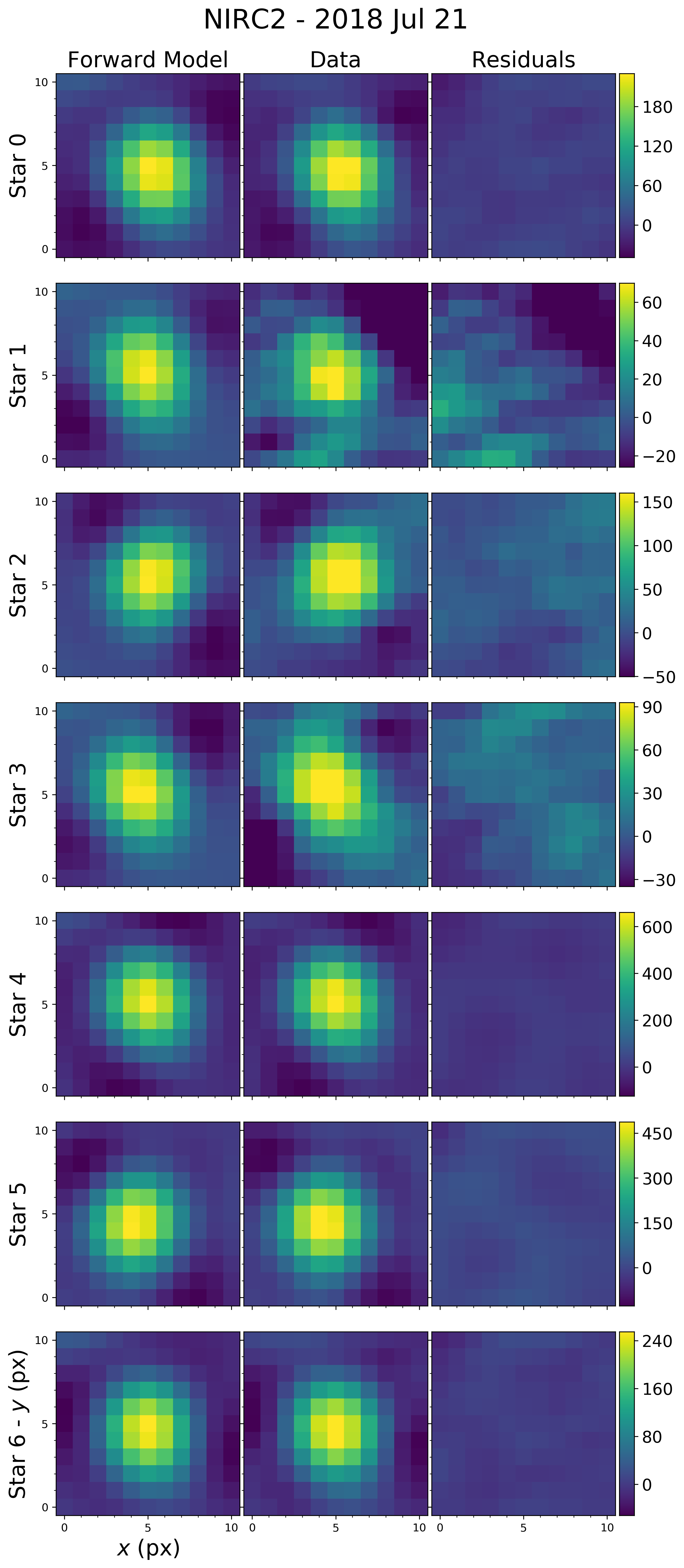}
\endminipage
\end{figure*}

\begin{figure*}
    \centering
\minipage{0.50\textwidth}
  \includegraphics[width=\linewidth]{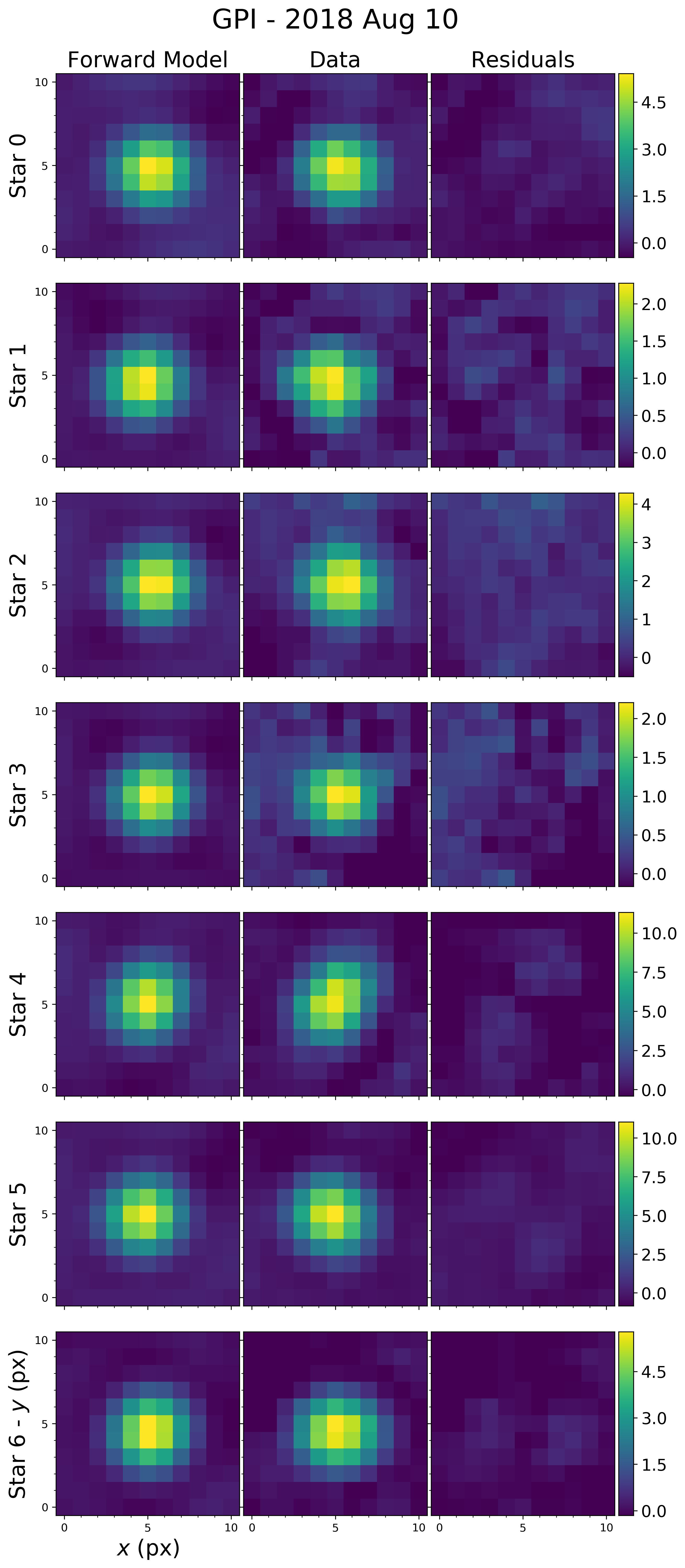}
\endminipage\hfill
\minipage{0.50\textwidth}
  \includegraphics[width=\linewidth]{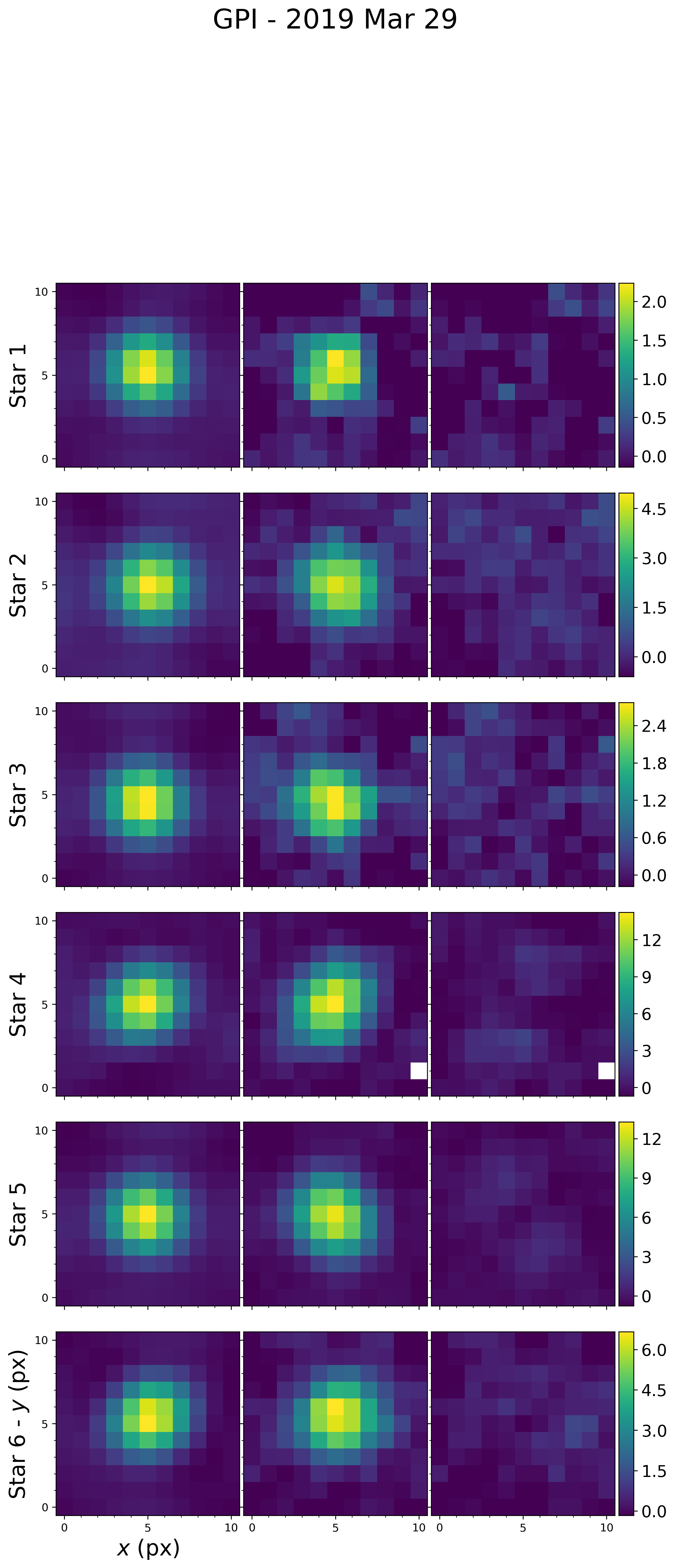}
\endminipage
\end{figure*}

\begin{figure*}
    \centering
\minipage{0.50\textwidth}
  \includegraphics[width=\linewidth]{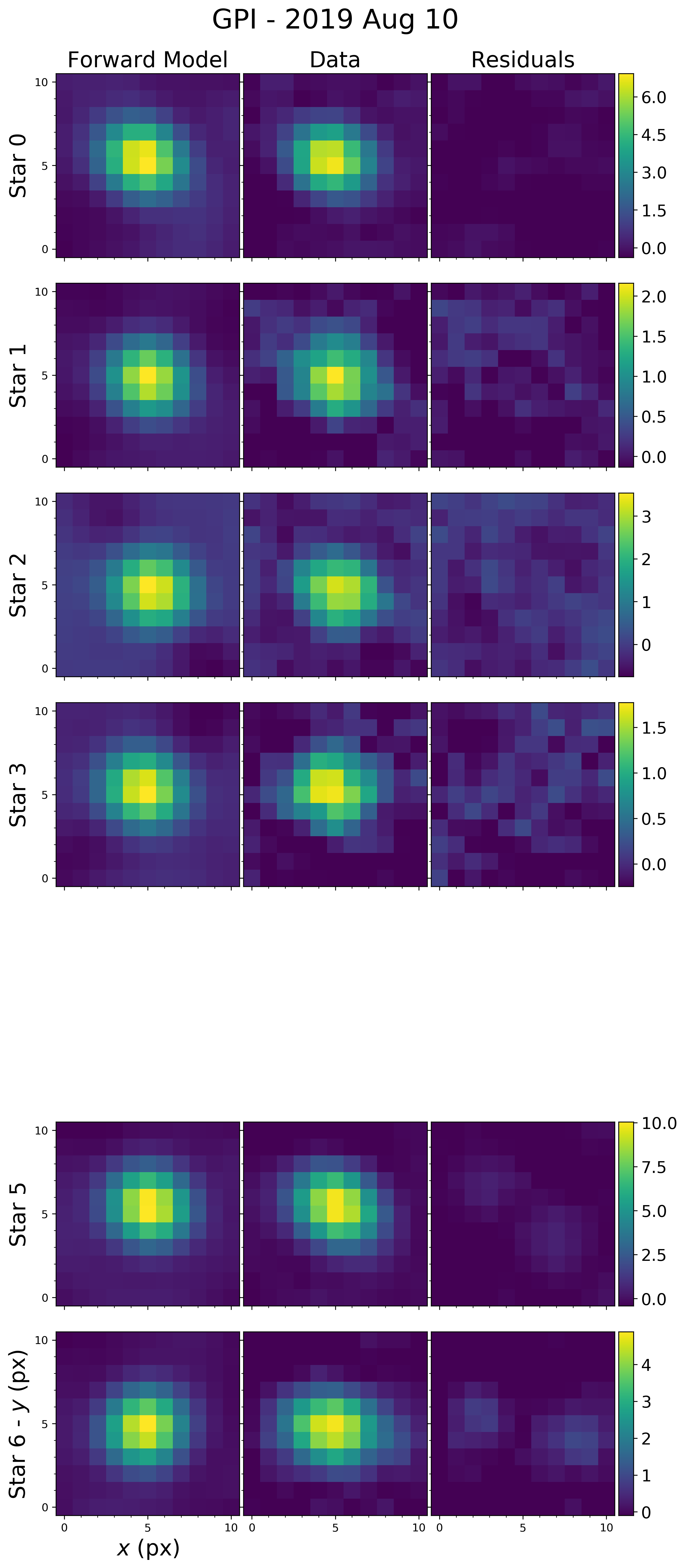}
\endminipage\hfill
\end{figure*}

% \figsetstart
% \figsetnum{11}
% \figsettitle{Stamp Gallery}

% \figsetgrpstart
% \figsetgrpnum{11.1}
% \figsetgrptitle{May 15, 2014}
% \figsetplot{stamp_gallery_140515.png}
% \figsetgrpend

% \figsetgrpstart
% \figsetgrpnum{11.2}
% \figsetgrptitle{Jul 12, 2014}
% \figsetplot{stamp_gallery_140712.png}
% \figsetgrpend

% \figsetgrpstart
% \figsetgrpnum{11.3}
% \figsetgrptitle{Jul 3, 2015}
% \figsetplot{stamp_gallery_150703.png}
% \figsetgrpend

% \figsetgrpstart
% \figsetgrpnum{11.4}
% \figsetgrptitle{Sep 1, 2015}
% \figsetplot{stamp_gallery_150901.png}
% \figsetgrpend

% \figsetgrpstart
% \figsetgrpnum{11.5}
% \figsetgrptitle{Apr 30, 2016}
% \figsetplot{stamp_gallery_160430.png}
% \figsetgrpend

% \figsetgrpstart
% \figsetgrpnum{11.6}
% \figsetgrptitle{Aug 7, 2017}
% \figsetplot{stamp_gallery_170807.png}
% \figsetgrpend

% \figsetgrpstart
% \figsetgrpnum{11.7}
% \figsetgrptitle{Jul 21, 2018}
% \figsetplot{stamp_gallery_180721.png}
% \figsetgrpend

% \figsetgrpstart
% \figsetgrpnum{11.8}
% \figsetgrptitle{Aug 10, 2018}
% \figsetplot{stamp_gallery_180810.png}
% \figsetgrpend

% \figsetgrpstart
% \figsetgrpnum{11.9}
% \figsetgrptitle{Mar 29, 2019}
% \figsetplot{stamp_gallery_190329.png}
% \figsetgrpend

% \figsetgrpstart
% \figsetgrpnum{11.10}
% \figsetgrptitle{Aug 10, 2019}
% \figsetplot{stamp_gallery_190810.png}
% \figsetgrpend

% \figsetgrpstart
% \figsetgrpnum{11.11}
% \figsetgrptitle{}
% \figsetplot{}
% \figsetgrpend

% \figsetend

% \begin{figure}
% \figurenum{11.1}
% \plotone{stamp_gallery_140515.png}
% \caption{Forward Model (Left), Data (Center), Residuals (Right)}
% \end{figure}

\section{Parallax Prior Considerations}

\begin{figure*}
\figurenum{12}
    \centering
    \includegraphics[width=\linewidth]{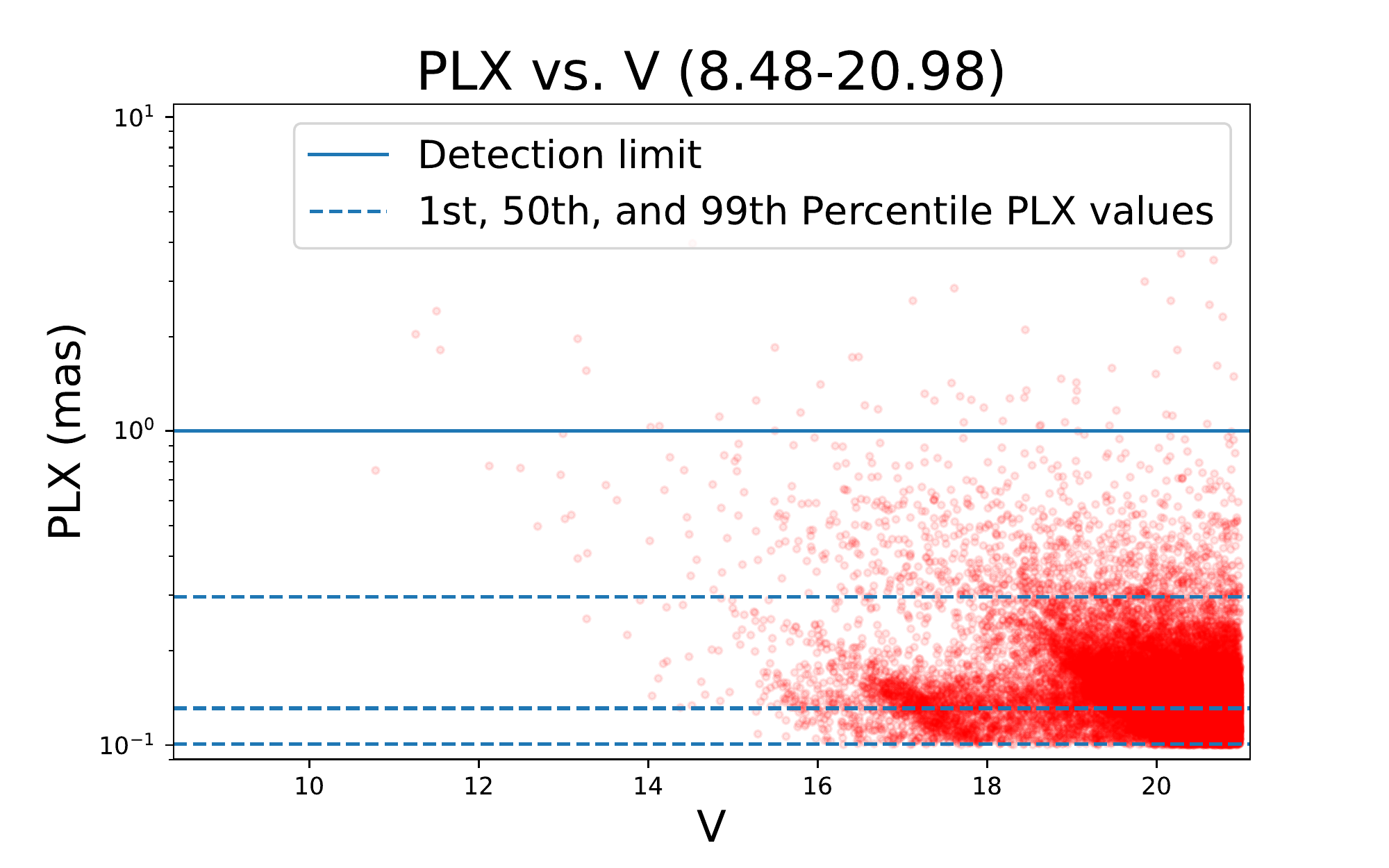}
    \caption{Distribution of Parallax vs. V-mag for stars within 12.5 V-mag of HD 165054 (8.48) which corresponds to GPI's theoretical maximum contrast difference of $\sim 10^{-5}$. As we can see, any functional dependence of parallax on V-mag is well below GPI's detection limit ($\sim 1 mas$).}
    \label{fig:plx_vs_V}
\end{figure*}

We include a figure of the distribution of Parallax vs. V-mag for stars within 12.5 V-mag of HD 165054 (8.48) for the purpose of assessing the flux independence of the background star parallaxes. See Fig. \ref{fig:plx_vs_V}.

\end{document}